\documentclass{optica-article}
\journal{opticajournal}
\articletype{Research Article}

\usepackage{amsmath}
\usepackage{graphicx}
\usepackage{lineno}
\usepackage{xcolor}
\usepackage{rotating}
\usepackage{subcaption}
\usepackage{tikz}
\usetikzlibrary{calc,positioning,fit,backgrounds,matrix,arrows.meta}
\usepackage{ifthen}
\usepackage{booktabs}
\usepackage{multirow}
\usepackage{siunitx}
\usepackage{adjustbox}
\definecolor{bestval}{RGB}{0,100,0}

\begin{document}

\title{A unified self-supervised framework for single-frame Fresnel CDI and overlapped ptychography}

\author{O. Hoidn,\authormark{1,*} S. Henke,\authormark{2} A. Vong,\authormark{2} A. Mishra,\authormark{1} A. Mehta,\authormark{1} and M. Seaberg\authormark{1}}
\address{\authormark{1}SLAC National Accelerator Laboratory, Menlo Park, California 94025, USA\\
\authormark{2}Argonne National Laboratory, Lemont, Illinois 60439, USA}
\email{\authormark{*}ohoidn@slac.stanford.edu}

\begin{abstract*}
Ptychographic imaging at synchrotron and X-ray free-electron laser sources requires densely overlapping scans, which limits throughput and increases dose; extending coherent diffractive imaging to overlap-free operation on extended samples remains an open problem. We present a self-supervised inverse-mapping network for single-frame Fresnel coherent diffraction imaging (CDI) and overlapped ptychography with fixed, pre-estimated probes. The learned neural network reconstructs individual object patches from either one diffraction frame or several overlapping measurements at a time. In single-frame mode, the phase diversity provided by the curved-wavefront probe at the off-focus sample position removes the requirement for overlap constraints, enabling sparser scans and proportionally lower dose at fixed exposure. On synthetic line patterns, reconstructed amplitude SSIM exceeds 0.90 in single-frame mode with the curved probe and reaches 0.952--0.968 with overlap constraints. Optimization of the network via a Poisson negative log likelihood objective, rather than the more common mean absolute error, yields $10$-fold improved photon-dose efficiency at doses below $10^5$ photons per image, where shot noise typically limits resolution. In addition to these synthetic studies, we demonstrate robust single-frame reconstruction of extended samples using ptychographic datasets from APS and LCLS, with end-to-end reconstruction of a 10,304-frame workload approximately $36\times$ faster than a highly optimized iterative solver. Together, these results unify single-frame Fresnel CDI and overlapped ptychography within one self-supervised framework, supporting dose-efficient, high-throughput imaging at modern light sources.

\end{abstract*}

\section{Introduction}\label{sec1}
Fourth-generation synchrotrons and X-ray free-electron lasers generate coherent diffraction data faster than conventional reconstruction pipelines can process them~\cite{LCLSIIHE_DesignPerf}. This growing gap between acquisition and analysis precludes real-time feedback and on-the-fly experimental steering, both essential for maximizing the scientific output of these facilities, and it likewise constrains dose-sensitive imaging.


Ptychographic coherent diffraction imaging (CDI) is a cornerstone X-ray nanoscale imaging technique~\cite{GuizarSicairos2021PhysicsToday,Pfeiffer2018XrayPtycho,Nugent2010AdvPhys,Chapman2010NatPhoton,Miao2015Science}, with demonstrations ranging from quantitative biological imaging and nanoscale tomography to integrated-circuit inspection and deep sub-{\aa}ngstr\"om electron microscopy~\cite{Giewekemeyer2010PNAS,Dierolf2010Nature,Holler2017Nature,Jiang2018Nature}. Single-frame Fresnel CDI and scanning ptychography constrain phase retrieval through different sources of measurement diversity. In Fresnel CDI, a known structured or defocused probe can encode sufficient phase information in a single diffraction pattern~\cite{Williams2006FresnelCDI,Abbey2008KeyholeCDI,Vine2009PFCDI}; some implementations generate this diversity using specialized optics such as randomized zone plates~\cite{Levitan2020Randomized}. Conventional reconstruction nevertheless remains iterative~\cite{GerchbergSaxton1972,Fienup1982,Marchesini2007Review,Shechtman2015PhaseRetrieval} and can be sensitive to initialization and support constraints~\cite{Williams2006FresnelCDI,Stockmar2013Nearfield,Marchesini2003Shrinkwrap}. Single-exposure coherent imaging is particularly attractive at free-electron lasers, where single-pulse diffraction can outrun radiation damage~\cite{Chapman2006FemtoCDI,Seibert2011Nature}; related multiple-shot schemes extend CDI to extended specimens~\cite{Takayama2021MultipleShotCDI}.

Ptychography~\cite{Hoppe1969,Rodenburg2004APL,Thibault2008Science} instead uses overlapping illumination footprints at neighboring scan positions, which allows relatively robust reconstruction using PIE-family or other iterative algorithms~\cite{Maiden2009UltramicroscopyPIE,GuizarSicairos2008TTD,Maiden2017rPIE}; robust convergence typically requires ${\sim}60$--$70\%$ scan overlap~\cite{Bunk2008Overlap,Edo2013PRA}, and analogous overlap-based reconstruction underlies visible-light Fourier ptychography~\cite{Zheng2013FPM}. Extensions jointly refine the probe~\cite{Thibault2009Probe} and correct scan-position errors~\cite{Maiden2012Annealing,Zhang2013Position}. Classical iterative solvers process only ${\sim}0.1$--$1$ diffraction patterns per second on standard hardware; a mature ecosystem of GPU-accelerated packages partially mitigates this limited throughput~\cite{Nashed2014Parallel,Marchesini2016SHARP,Enders2016PtyPy,FavreNicolin2020PyNX,Wakonig2020PtychoShelves,Babu2023EdgePtycho}, but throughput remains a practical obstacle to real-time imaging.

Trained neural networks amortize the cost of inversion: an approximate inverse map from reciprocal space to real space is learned once from many training samples, after which the model supports rapid single-pass inference~\cite{Sinha2017Optica,Rivenson2018LSA,Barbastathis2019Optica,Cherukara2018SciRep}. However, the training procedure typically requires paired object labels and therefore the use of iterative solvers to generate a training corpus, and the resulting models often generalize poorly beyond their training distribution~\cite{Cherukara2020PtychoNN,Babu2023EdgePtycho}. In addition, the lack of physics-consistency constraints in such models tends to limit their accuracy~\cite{Hoidn2023PtychoPINN}. In short, neither conventional iterative methods nor direct supervised inversion unifies speed, resolution, and flexible handling of real-space constraints.

Beyond supervised direct inversion, recent machine learning approaches to ptychographic phase retrieval include hybrid physics--learning methods such as deep-prior regularization~\cite{Ulyanov2018DIP,Metzler2018prDeep} and learned accelerators within iterative solvers~\cite{McCray2025IntegratedNN}, implicit neural representations including SIREN-style parameterizations~\cite{Sitzmann2020SIREN}, learned probe-position correction for large scan errors~\cite{Du2024ProbePosition}, unrolled transformer-based ptychography networks~\cite{Gan2024PtychoDV}, and deep networks for single-shot ptychography~\cite{Wengrowicz2020DeepSSP}. Related physics-aware and unsupervised inversion has also been demonstrated for 3D and Bragg CDI~\cite{Chan2021APR,Yao2022AutoPhaseNN}. Within ptychography and extended-sample CDI, however, no single prior approach has jointly demonstrated reusable pre-trained inference, label-free training, and operation without neighboring-frame overlap constraints.

Existing trained physics-based Fourier phase retrieval techniques improve on purely data-driven approaches by using the diffraction forward model to generate synthetic training data and incorporating image-formation consistency in fine-tuning~\cite{Zhang2024FPR}. Our previously introduced PtychoPINN further reduces the requirement for real-space ground truth information by fitting an inverse-mapping neural network via a differentiable coherent-scattering forward model incorporated within the network; in training, it dispenses entirely with real-space supervision and uses only agreement between predicted and measured diffraction as the training signal. This measurement-consistency objective parallels the broader physics-informed learning paradigm~\cite{Raissi2019PINN,Karniadakis2021PIML}.


Our earlier work introduced PtychoPINN for synthetic, grid-sampled ptychography, where overlapping measurements jointly constrained the reconstruction~\cite{Hoidn2023PtychoPINN}. Here, we extend the same inverse--forward formulation to single-frame Fresnel CDI, treating real-space redundancy as a configurable parameter rather than a hard requirement: when a curved or defocused probe provides sufficient phase diversity, the diffraction-domain likelihood alone can anchor the reconstruction---the principle underlying Fresnel CDI. In single-frame mode, the network infers each object patch from one diffraction pattern using a fixed, pre-estimated probe, without coupling the reconstruction to neighboring measurements; supplied scan coordinates are then used to place the recovered patches within an extended object. When overlapping frames are reconstructed jointly, the same formulation additionally enforces consistency across their shared object region. Separating single-frame recovery from coordinate-based assembly thus removes neighboring-frame overlap as an acquisition requirement, permitting a fixed field of view to be sampled with fewer exposures, and therefore at lower total dose for a fixed exposure per frame. Throughout, we use ``single-frame'' in the restricted sense of one diffraction measurement with a structured probe, as distinct from single-shot schemes based on beam multiplexing~\cite{Sidorenko2015Optica,Kharitonov2022SciRep} or downstream modulators~\cite{Zhang2016CMI,Dong2018CMI}.

We evaluate this extension using complementary synthetic and experimental studies. Synthetic experiments compare single-frame and multi-frame reconstruction across probe structures, benchmark PtychoPINN against a distinct supervised model, and isolate the effect of loss-function choice under photon-limited conditions. Experimental studies evaluate APS (Velociprobe/2-ID) and LCLS (XPP) reconstructions both within their respective data distributions and under APS-to-LCLS application without retraining. We also compare the end-to-end reconstruction time of the trained framework to that of highly optimized iterative solvers. Specifically, we demonstrate (i) self-supervised single-frame reconstruction of experimental APS and LCLS data, with feedforward inference at ${\sim}6.1\times10^3$ diffraction patterns per second at $64\times64$ and an approximately $36\times$ end-to-end reconstruction-time advantage over an optimized iterative solver; (ii) overlap-free single-frame reconstruction in a Fresnel CDI geometry; (iii) dose-efficient imaging via a Poisson likelihood objective, which matches the resolution of amplitude-MAE training with tenfold fewer simulated photons per frame; and (iv) roughly an order-of-magnitude improvement in training-data efficiency over a distinct supervised baseline.

\section{Methods and Architecture}
\label{sec:methods}

\subsection{Formulation and Forward Model}
\label{sec:formulation}

We learn an inverse map $G: X \!\to\! Y$ from diffraction space to real space and optimize it by composing with a differentiable forward model $F: Y \!\to\! X$. The autoencoder $F \circ G$ is trained to match measured diffraction statistics without ground-truth images.

\paragraph{Data model and notation.}
Each training example comprises $C_g$ nonnegative real detector-amplitude arrays $a'_k$ of shape $N\times N$, acquired at supplied absolute probe positions $\vec r_k$, with $k=1,\ldots,C_g$ and $N$ the detector/patch side length. For count-resolved data, $a'_{kij}=\sqrt{n'_{kij}}$, where $n'_{kij}$ is the measured count at detector pixel $(i,j)$; otherwise $a'_k$ is the stored measured-amplitude array in its stated scale. The inverse map produces $N\times N$ complex object patches $O_k$. Table~\ref{tab:symbols} in Appendix~A summarizes the notation.

We use $\vec r$ for an object-plane grid coordinate and $M$ for the merged-canvas side length. The operator $\mathcal T_{\Delta\vec r}$ translates a finite-grid image, $\mathrm{Pad}_M$ zero-pads an $N\times N$ patch to the canvas, and $\mathrm{Crop}_N$ takes the central $N\times N$ crop. The support mask $M_{\mathrm{center}}$ is one on the central $N/2\times N/2$ tile and zero elsewhere.
\paragraph{Probe overlap and grouping.}
The \emph{joint reconstruction group size} $C_g$ is the number of frames decoded, registered, and merged within one example. \emph{Single-frame reconstruction/inference} is the special case $C_g=1$, in which one detector frame produces one object patch. A densely sampled scan may provide many independent $C_g=1$ examples without introducing a neighboring-frame overlap relation into the reconstruction procedure.

\paragraph{Constraint map ($F_c$): translation-aware merging.}
With $F=F_d\circ F_c$, the same shift-and-normalize constraint map $F_c$ applies for every $C_g$; $F_c$ couples distinct decoded patches only when $C_g>1$:
\begin{align}
  O_{\mathrm{region}}(\vec r)
  &=
  \frac{
    \displaystyle\sum_{k=1}^{C_g}
    \mathcal T_{-\vec r^{\,\mathrm{rel}}_k}
    \!\left[\mathrm{Pad}_M(O_k)\right]
  }{
    \displaystyle\sum_{k=1}^{C_g}
    \mathcal T_{-\vec r^{\,\mathrm{rel}}_k}
    \!\left[\mathrm{Pad}_M(M_{\mathrm{center}})\right]
    +\epsilon
  },
  \qquad \epsilon=10^{-3}.
  \label{eq:constraintmap}
\end{align}
Here $O_{\mathrm{region}}$ is the $M\times M$ complex merged-object representation, and the denominator counts the translated central supports; the fixed $\epsilon$ is a numerical safeguard that prevents division by zero outside the accumulated support, not a fitted reconstruction parameter. For non-integer offsets, $\mathcal T$ is a bilinear subpixel warp with zero fill outside the canvas; because interpolation, finite support, padding, and cropping can discard energy, this discrete translation is not unitary. This ``translational pooling'' applies to arbitrary scan geometries.

\paragraph{Coordinate-aware grouping.}
Source-frame centers are selected from the supplied scan. For $C_g>1$, each example is formed from a local $K$-nearest-position candidate pool, where $K$ is the configured neighbor count ($K=4$ by default). The requested number of training groups is separate from both $K$ and the number of source frames.

Concretely, for each anchor position $\vec r_i$, let $\mathcal N_K(\vec r_i)$ denote its $K$ nearest distinct scan positions. A group contains the anchor together with $C_g-1$ neighbors sampled uniformly without replacement from $\mathcal N_K(\vec r_i)$, repeated $n_{\mathrm{samples}}$ times per anchor. If duplicate neighbor sets are disallowed, the number of distinct groups per anchor is
\begin{equation}
  n_{\mathrm{grp}}
  =\min\!\left(n_{\mathrm{samples}},\,\binom{K}{C_g-1}\right),
  \label{eq:ngroups}
\end{equation}
so a scan of $N_{\mathrm{scan}}$ positions yields up to $N_{\mathrm{scan}}\,n_{\mathrm{grp}}$ training examples. Choosing $n_{\mathrm{samples}}>1$ augments the dataset through combinatorial re-grouping while preserving local spatial consistency.

For the $C_g$ coordinates actually placed in a group, define the centroid and the stored model offsets by
\begin{equation}
  \bar{\vec r}
  =\frac{1}{C_g}\sum_{k=1}^{C_g}\vec r_k,
  \qquad
  \vec r^{\,\mathrm{rel}}_k
  =\bar{\vec r}-\vec r_k .
  \label{eq:relativecoords}
\end{equation}
The sign matters: the constraint map translates decoded patches by $-\vec r^{\,\mathrm{rel}}_k$, whereas the diffraction map below extracts them with $+\vec r^{\,\mathrm{rel}}_k$. When $C_g=1$, the group contains the selected frame itself, $\bar{\vec r}=\vec r_1$, and $\vec r^{\,\mathrm{rel}}_1=\vec 0$. Equation~\eqref{eq:constraintmap} then has one term and supplies no pairwise overlap constraint; the diffraction loss is evaluated for that frame alone. Additional scan frames increase the number of independent examples, not the number of frames or coordinate-consistency terms inside a $C_g=1$ example.

\paragraph{Diffraction map ($F_d$): coherent scattering.}
Given the merged object, the forward map extracts each object patch, multiplies it by the fixed effective probe, and predicts detector amplitude and intensity:
\begin{align}
  O'_k(\vec r)
  &=\mathrm{Crop}_N\!\left[
    \mathcal T_{\vec r^{\,\mathrm{rel}}_k}(O_{\mathrm{region}})
  \right],\\
  \psi_k(\vec r)
  &=O'_k(\vec r)P(\vec r),\qquad
  \Psi_k=\operatorname{FFT2}_N\!\{\psi_k\},\\
  A^{(0)}_{kij}
  &=\frac{1}{N}
    \left[\operatorname{fftshift}\!\left(|\Psi_k|^2\right)_{ij}\right]^{1/2},\\
  \hat A_{kij}
  &=s_\alpha A^{(0)}_{kij},\qquad
  \mu_{kij}=\hat A_{kij}^{\,2}
  =s_\alpha^2\!\left(A^{(0)}_{kij}\right)^2,
  \qquad s_\alpha\equiv e^{\alpha_{\log}} .
  \label{eq:forward}
\end{align}
Here $O'_k$ is the $N\times N$ extracted complex object patch, $\psi_k$ the exit wave, and $\Psi_k$ its unnormalized discrete Fourier transform; $A^{(0)}_k$ is the internally normalized detector amplitude, $\hat A_k$ the target-scale predicted amplitude, and $\mu_k$ the predicted Poisson rate. The discrete implementation evaluates an unpadded $N\times N$ FFT and divides $|\Psi_k|^2$ by $N^2$ before taking the square root, reproducing the $1/N$ amplitude normalization above. No second Fourier normalization is used. The shared $s_\alpha$ is therefore an amplitude scale, and $s_\alpha^2$ is its intensity-scale counterpart.

The use of one Fourier transform for the Fresnel-regime data follows the Fresnel scaling theorem~\cite{Paganin2006CoherentXRayOptics,Williams2006FresnelCDI}. Let $\vec x$ and $\vec u$ be sample- and detector-plane coordinates, $\lambda_{\mathrm{x}}$ the X-ray wavelength, $z=d_2$ the sample--detector distance (Fig.~\ref{fig:setup} and Table~\ref{tab:setup_params}), and
$C_z(\vec x)=\exp[\mathrm{i}\pi|\vec x|^2/(\lambda_{\mathrm{x}}z)]$, where $\mathrm{i}^2=-1$. After dropping unit-modulus output-phase factors that cancel in intensity, the detector intensity $I_z$ under scalar paraxial propagation is
\begin{equation}
  I_z(\vec u)
  =\frac{1}{(\lambda_{\mathrm{x}}z)^2}
  \left|
    \mathcal F\!\left\{
      P_{\mathrm{physical}}(\vec x)O(\vec x)C_z(\vec x)
    \right\}
    \!\left(\frac{\vec u}{\lambda_{\mathrm{x}}z}\right)
  \right|^2 .
  \label{eq:fresnelscaling}
\end{equation}
Here $\mathcal F$ is the continuous two-dimensional Fourier transform, $P_{\mathrm{physical}}$ is the physical sample-plane probe, $O$ is the thin-sample complex transmission, and $C_z^*$ denotes the complex conjugate of $C_z$. Thus $P_{\mathrm{eff}}=P_{\mathrm{physical}}C_z$. Under the single-Fourier retrieval convention used here, the fixed probe in Eq.~\eqref{eq:forward} is $P=P_{\mathrm{rec}}\equiv P_{\mathrm{eff}}$, or equivalently $P_{\mathrm{physical}}=P_{\mathrm{rec}}C_z^*$; no second chirp is applied in the network forward map.

This representation assumes a scalar, monochromatic, coherent paraxial field in homogeneous free space, a thin multiplicative sample, and probe and object defined in the same plane. On the discrete grids, the sample pitch $\Delta x$ and detector pitch $\Delta u$ satisfy $\Delta x\,\Delta u=\lambda_{\mathrm{x}}z/N$, with centered grids and adequate sampling and support for $C_z$. For LCLS XPP, $\lambda_{\mathrm{x}}=1.3947576$~\AA{}, $z=4.147$~m, $\Delta u=75$~$\mu$m, and $N=64$ give $\Delta x=120.5$~nm. Constant propagation and transform prefactors are absorbed into the recorded amplitude scale. These conditions exclude, without further modeling, thick or multiple-scattering samples, partial coherence or appreciable bandwidth, and inconsistent or undersampled geometry. Supplementary Sec.~S1 and Fig.~S1 verify the numerical equivalence on the LCLS XPP grid; they do not independently validate the experimental geometry.

\subsection{Data Preprocessing}
\label{sec:preprocess}

A dataset contains detector frames from one or more objects measured with probe illumination $P$. In the reported experiments, conventional iterative ptychography with probe refinement~\cite{Thibault2009Probe,Maiden2009UltramicroscopyPIE} supplies $P_{\mathrm{rec}}$, which is then held fixed as $P$ in the differentiable forward model; neither the probe nor the supplied positions are refined during object-network training. The probe is a forward-model parameter, not a real-space supervision target.

For LCLS XPP, events were retained when the intensity-position-monitor (IPM) reading satisfied $40000<\texttt{ipm2}<50000$ and the corresponding measured scan coordinates were finite. A $64\times64$ detector ROI centered at detector coordinates $(150,150)$ pixels was extracted from \texttt{jungfrau1M.ROI\_0\_area}. The archived detector-processing workflow set intensities below 20 to zero, divided each retained frame by \texttt{ipm2}, and took the square root to obtain the normalized diffraction amplitudes supplied to the extended ptychographic iterative engine (ePIE) and the neural-network workflow. The final dataset contained 1,087 accepted frames. The displayed and reconstructed LCLS arrays are therefore measured, IPM-normalized diffraction amplitudes, not unprocessed raw detector counts.

After frames are grouped, measured detector amplitudes are normalized for order-unity network activations:
\begin{align}
  s_{\mathrm{data}}
  &=
  \frac{
    \sqrt{\left\langle\sum_{i,j}|a'_{kij}|^2\right\rangle_k}
  }{N/2},
  &
  x_{kij}
  &=\frac{a'_{kij}}{s_{\mathrm{data}}}.
  \label{eq:norm}
\end{align}
The average $\langle\cdot\rangle_k$ is over all detector frames in the grouped dataset being normalized, so
$\left\langle\sum_{i,j}|x_{kij}|^2\right\rangle_k=N^2/4$. This choice yields order-unity network activations: by Parseval's theorem, a unit-amplitude object confined to the central $N/2\times N/2$ support produces diffraction power of approximately $N^2/4$, so the normalization maps measured amplitudes onto the scale of unit-amplitude objects. For count-resolved data, this preserves $a'=\sqrt{n'}$; for amplitude-only records, it acts on the stored amplitude scale as given.

The arrays presented to the compiled model and its losses are fixed before optimization:
\begin{equation}
  a_k=s_0x_k,\qquad n_k=a_k^2,\qquad
  \widetilde x_k=\frac{a_k}{s_\alpha}
  =\frac{s_0}{s_\alpha}x_k .
  \label{eq:internalscale}
\end{equation}
Here $s_0$ is the fixed dataset-to-target amplitude factor, $a_k$ and $n_k$ are the external amplitude and intensity targets, and $s_\alpha=e^{\alpha_{\log}}$ is the shared internal model scale with log-amplitude parameter $\alpha_{\log}$; $s_\alpha$ divides the model input and multiplies $A_k^{(0)}$ in Eq.~\eqref{eq:forward}. Initially $s_\alpha(0)=s_0$, so $G$ receives $\widetilde x_k=x_k$. If $s_\alpha$ is trainable, the internal input and prediction scales change together, while the external targets $a_k$ and $n_k$ remain fixed.

\subsection{Neural Network Architecture}
\label{sec:nn}

The inverse map $G$ is an encoder--decoder (as in \cite{Hoidn2023PtychoPINN}; see also \cite{Vong2025GeneralizablePtycho} for a related PyTorch implementation). It maps the grouped detector amplitudes $\{\widetilde x_k\}_{k=1}^{C_g}$ to complex object patches $\{O_k\}_{k=1}^{C_g}$; supplied scan offsets enter only the fixed translation and extraction operators, not the learned network.

Restricting the high-resolution reconstruction to the central $N/2\times N/2$ tile satisfies the oversampling condition~\cite{miao1999extending}, but probes with extended tails use this limited support inefficiently: the brightly illuminated region is small compared with the total probe area that must be represented to avoid truncation artifacts from nonzero amplitude at the edge of the real-space grid. Fully inscribing the probe, tails included, within the central tile can therefore require coarse binning, forcing a choice between truncation artifacts (and the associated physical inconsistency) and violation of the oversampling condition. We resolve this by reconstructing the object at high resolution in the central tile and at low resolution in the periphery; provided the probe tail lacks high spatial frequencies, the peripheral continuation does not compromise well-posedness of the inverse problem.

Concretely, the decoder uses bounded amplitude and phase heads for the central $N/2\times N/2$ tile and four low-capacity channels for the coarse peripheral continuation that reduces probe/exit-wave truncation beyond the conventional oversampled support. The resulting patch is $O_k=O_{\mathrm{amp}}\exp(\mathrm{i}O_{\mathrm{phase}})$; the central amplitude and phase lie in $(0,1)$ and $(-\pi,\pi)$, while the peripheral continuation uses unbounded swish activations only outside the central tile.

\subsection{Training Objective and Optimization}
\label{sec:loss}

Training optimizes the inverse map $G$ through diffraction-domain objectives: a detector-amplitude mean absolute error (MAE) and a Poisson negative log likelihood (NLL) on detector intensity~\cite{Thibault2012NJPML,Seifert2023PoissonGaussian}. With batch size $B$, group size $C_g$, and $N\times N$ detector arrays,
\begin{align}
  \mathcal L_{\mathrm{MAE}}
  &=
  \frac{1}{B C_g N^2}
  \sum_{b,k,i,j}
  \left|\hat A_{bkij}-a_{bkij}\right|,\\
  \mathcal L_{\mathrm{Poiss}}
  &=
  \frac{1}{B C_g N^2}
  \sum_{b,k,i,j}
  \left(\mu_{bkij}-n_{bkij}\log\mu_{bkij}\right),
  \qquad
  \mu_{bkij}=\hat A_{bkij}^{\,2},\quad
  n_{bkij}=a_{bkij}^{\,2}.
  \label{eq:poissonloss}
\end{align}
The omitted $\log\Gamma(n+1)$ term is independent of the prediction. Training uses
\begin{equation}
  \mathcal L
  =w_{\mathrm{MAE}}\mathcal L_{\mathrm{MAE}}
  +w_{\mathrm{Poiss}}\mathcal L_{\mathrm{Poiss}},
\end{equation}
with the real-space supervision weight fixed to zero for PtychoPINN.

Let $N_{\mathrm{eff}}$ denote the configured mean integrated target intensity per frame. The fixed target scale and initial internal amplitude scale are
\begin{equation}
  s_0\equiv s_\alpha(0)
  =\frac{2\sqrt{N_{\mathrm{eff}}}}{N}.
  \label{eq:alphaloginit}
\end{equation}
Together with Eq.~\eqref{eq:norm}, this gives
$\left\langle\sum_{i,j}n_{kij}\right\rangle_k=N_{\mathrm{eff}}$.
The shared $s_\alpha$ may remain fixed or be trained; learning it can absorb modest calibration errors in the imposed intensity scale. The external targets and $s_0$ remain fixed. Equation~\eqref{eq:poissonloss} is a calibrated likelihood only when the target intensities retain a detector-count interpretation.

The APS and LCLS models use $(w_{\mathrm{MAE}},w_{\mathrm{Poiss}})=(0,1)$, $N_{\mathrm{eff}}=10^9$, and trainable $s_\alpha$. Their detector inputs are normalized and rescaled amplitudes, so the experimental loss is a Poisson-form intensity discrepancy at an imposed scale rather than a calibrated photon-count likelihood. The controlled overlap ablation uses amplitude MAE, while the synthetic photon-count study directly compares amplitude MAE with count-calibrated Poisson NLL.

\paragraph{Implementation notes.}
The TensorFlow graph composes bilinear translations, zero-padding and center-cropping, fixed-probe multiplication, an unpadded FFT, the shared amplitude scale, and the selected diffraction-domain objective. These operations are differentiable with respect to the decoded object and, when enabled, $s_\alpha$; supplied coordinates and the fixed probe are not optimized. At $C_g=1$ the cross-frame merge branch is absent; at $C_g>1$ neighboring decoded patches are translated and merged by Eq.~\eqref{eq:constraintmap}. Differentiable diffraction forward models built on automatic differentiation have likewise been used to generalize iterative ptychographic and tomographic reconstruction~\cite{Kandel2019AD,Du2021Adorym}. Appendix~A (Table~\ref{tab:config_params}) lists the model, training, and evaluation settings.

\subsection{Baselines and Reference Reconstructions}
The supervised baseline is a distinct PtychoNN-style dual-head encoder--decoder~\cite{Cherukara2020PtychoNN}. Each example contains one normalized diffraction-amplitude frame ($C_g=1$); the network receives no scan-coordinate input. A shared convolutional encoder feeds separate amplitude and phase decoders, which are optimized directly against paired real-space amplitude and phase target arrays using mean absolute error. The supervised baseline therefore contains neither the differentiable diffraction operator nor a detector-domain forward-model loss. In learned-model comparisons, this supervised model uses the same training frame budget and the same spatial split as the corresponding PtychoPINN model. Simulated-object targets are known by construction. For LCLS XPP, supervised target patches are extracted from the archived ePIE object reconstruction, and the ePIE reference array used for evaluation is byte-identical to that reconstruction. Although the held-out coordinates do not overlap the training coordinates, the supervised model was trained to reproduce the same ePIE reconstruction that serves as the evaluation reference, so comparison against that reference may favor it.

The two iterative reconstructions serve different roles. The Tike rPIE~\cite{Maiden2017rPIE} reconstruction is a shared full-data APS reference generated from all 10{,}304 frames with 1{,}000 iterations; it is not matched to the 512- or 5{,}050-frame learned-model budgets.

The LCLS XPP ePIE reconstruction~\cite{Maiden2009UltramicroscopyPIE} initializes the object as a complex array of ones and the probe with Gaussian amplitude and quadratic phase, normalized to the mean diffraction power. Object and probe feedback parameters are $\alpha=\beta=1$, and the scan positions are visited in a newly randomized order during every pass; the random seed was not recorded. In stage~1, ePIE runs 30 passes, updates the probe during passes 0--9, resets the object before passes 10 and 20, and performs no position correction. In stage~2, it runs 60 passes, performs position correction during passes 1--29 with $\gamma$ initialized to 50 and then adjusted dynamically by the reconstruction, updates the probe during passes 30--39, and resets the object before pass 40. The probe and scan positions are held fixed during passes 40--59. This same-acquisition ePIE reconstruction is a reference rather than ground truth; because its object supplied the supervised training targets, comparison with ePIE may favor the supervised model.

\subsection{Experimental Setup and System Parameters}

The APS 2-ID Velociprobe, as described in the cited publications, uses a Fresnel zone plate, and the LCLS XPP instrument~\cite{Chollet2015XPP} uses a beryllium compound-refractive-lens stack; each optic forms a focus upstream of the sample. The APS object-plane sampling is 60.3~nm\,px$^{-1}$, while the LCLS XPP energy, wavelength, detector geometry, and distance give 120.5~nm\,px$^{-1}$. In both geometries, the exit wave propagates without intervening optics from the sample to the detector.

\begin{figure}[htbp]
  \centering
  \begin{tikzpicture}[
    x=0.92cm,
    y=1cm,
    font=\footnotesize,
    >=Stealth,
    ray/.style={draw=black!85, line width=0.65pt, line cap=round},
    optic/.style={draw=black!85, fill=black!10, line width=0.65pt},
    dimension/.style={<->, draw=black!85, line width=0.55pt},
    provenance/.style={draw=black!55, fill=black!4, rounded corners=1pt,
      inner xsep=3pt, inner ysep=2pt, font=\scriptsize\bfseries}
  ]
    \def\xOptic{0.55}
    \def\xFocus{2.35}
    \def\xSample{4.25}
    \def\xDetector{7.75}

    \node[provenance, anchor=west] at (0,2.25) {APS 2-ID Velociprobe};
    \node[provenance, anchor=west] at (0,1.86) {LCLS XPP};

    \draw[optic] (\xOptic-0.08,-0.58) rectangle (\xOptic+0.08,0.58);
    \foreach \yy in {-0.40,-0.20,0,0.20,0.40}
      \draw[black!55, line width=0.35pt] (\xOptic-0.08,\yy) -- (\xOptic+0.08,\yy);
    \node[align=center] at (\xOptic,-0.92) {focusing optic\\(FZP / Be lens stack)};

    \draw[ray] (\xOptic+0.08,0.55) -- (\xFocus,0);
    \draw[ray] (\xOptic+0.08,-0.55) -- (\xFocus,0);

    \fill[black!85] (\xFocus,0) circle (1.4pt);
    \node at (\xFocus,-0.34) {focus};

    \draw[ray] (\xFocus,0) -- (\xSample,0.50);
    \draw[ray] (\xFocus,0) -- (\xSample,-0.50);
    \draw[optic, fill=black!22]
      (\xSample-0.045,-0.53) rectangle (\xSample+0.045,0.53);
    \node at (\xSample,0.76) {sample};

    \draw[ray] (\xSample,0.50) -- (\xDetector-0.15,0.88);
    \draw[ray] (\xSample,-0.50) -- (\xDetector-0.15,-0.88);
    \draw[ray] (\xSample,0) -- (\xDetector-0.15,0);
    \draw[optic, fill=black!5]
      (\xDetector-0.15,-0.90) rectangle (\xDetector+0.15,0.90);
    \foreach \yy in {-0.60,-0.30,0,0.30,0.60}
      \draw[black!45, line width=0.35pt]
        (\xDetector-0.15,\yy) -- (\xDetector+0.15,\yy);
    \node at (\xDetector,1.16) {detector};
    \node[font=\scriptsize\itshape, text=black!70] at (6.00,0.30)
      {free-space propagation};

    \draw[dimension] (\xFocus,-1.36) -- (\xSample,-1.36)
      node[midway, below] {$d_1$ (defocus)};
    \draw[dimension] (\xSample,-1.36) -- (\xDetector,-1.36)
      node[midway, below] {$d_2$};
  \end{tikzpicture}
  \caption{Experimental setup schematic (not to scale). The focusing optic is a Fresnel zone plate (FZP) at APS 2-ID and a beryllium compound-refractive-lens stack at LCLS XPP.}
  \label{fig:setup}
\end{figure}

\begin{table}[htbp]
  \centering
  \caption{Experimental setup, acquisition, and reconstruction-grid parameters. $\dagger$ marks cited nominal APS geometry supplied for context. Physical dimensions in the figures use the calibrated reconstruction-grid samplings rather than values inferred from the nominal geometry.}
  \label{tab:setup_params}
  \begin{adjustbox}{max width=\textwidth}
  \footnotesize
  \renewcommand{\arraystretch}{1.14}
  \begin{tabular}{@{}p{0.24\textwidth}p{0.35\textwidth}p{0.35\textwidth}@{}}
    \toprule
    Parameter & APS 2-ID Velociprobe (fly64) & LCLS XPP \\
    \midrule
    \multicolumn{3}{@{}l}{\textit{Dataset and reconstruction arrays}} \\
    Diffraction frames & $10{,}304$ & $1{,}087$ accepted frames \\
    Analyzed detector region & $64\times64$ pixels & Jungfrau 1M ROI; $64\times64$ pixels centered at detector coordinates $(150,150)$ pixels \\
    Reconstruction input & Raw detector counts & Measured, IPM-normalized diffraction amplitudes \\
    Probe & $64\times64$ complex array & $64\times64$ complex array \\
    \midrule
    \multicolumn{3}{@{}l}{\textit{Physical setup}} \\
    Photon energy (wavelength) & $8.8$\,keV ($1.41$\,\AA{}) $\dagger$~\cite{Deng2022MultiPass} & $8889$\,eV ($1.3947576$\,\AA{}) \\
    Focusing optic & FZP, $50$\,nm outer zone $\dagger$~\cite{Deng2019Velociprobe} & Beryllium compound-refractive-lens stack \\
    Focus--sample ($d_1$) & $\approx600\,\mu$m $\dagger$~\cite{Deng2022MultiPass} & $1.00$\,mm \\
    Sample--detector ($d_2$) & $1.92$\,m $\dagger$~\cite{Deng2022MultiPass} & $4.147$\,m \\
    Detector pixel pitch & Eiger X 500K, $75\,\mu$m $\dagger$~\cite{Deng2022MultiPass} & Jungfrau 1M ROI, $75\,\mu$m \\
    Native detector array & $1028\times512$ pixels $\dagger$~\cite{Deng2022MultiPass} & --- \\
    Scan coordinates & $100$\,nm nominal step $\dagger$~\cite{Deng2022MultiPass} & Measured per-frame piezo coordinates supplied directly \\
    Object-plane sampling & $60.3$\,nm\,px$^{-1}$ & $120.5$\,nm\,px$^{-1}$ \\
    $64$-pixel field of view & $3.86$\,$\mu$m & $7.71$\,$\mu$m \\
    \bottomrule
  \end{tabular}
  \end{adjustbox}
\end{table}

\subsection{Datasets and Evaluation Protocol}
We evaluate APS Velociprobe Siemens-star measurements, LCLS XPP test-pattern measurements, a synthetic Siemens-star dataset simulated from APS reconstructions, and the synthetic line-pattern dataset of Ref.~\cite{Hoidn2023PtychoPINN}. All learned-model reconstructions of the APS and LCLS data use single-frame examples ($C_g=1$); the synthetic line-pattern ablation additionally evaluates four-frame groups ($C_g=4$) formed from adjacent, overlapping scan positions. The APS experiments use a low-$y$/high-$y$ spatial split; models are trained on either all 5,050 training-half frames or a random 512-frame subsample of them, and evaluated over the full scan. The synthetic Siemens-star studies use known simulated-object ground truth. For cross-facility evaluation, we apply the APS-trained models to LCLS XPP diffraction amplitudes with the learned network and its weights unchanged, substituting only the beamline-specific inference inputs (LCLS probe, scan coordinates, and object-plane sampling) at evaluation. The learned models receive diffraction amplitudes but no coordinates; stored scan positions enter fixed PtychoPINN operators and external full-field assembly, with no position refinement.

Where ground truth or a reference amplitude is available, we report structural similarity (SSIM). The line-pattern SSIM is computed from amplitude. In the cross-facility analysis, amplitude MSE and peak signal-to-noise ratio (PSNR), $\mathrm{PSNR}=10\log_{10}(R^2/\mathrm{MSE})$, use the unscaled post-trim reconstruction and ePIE-reference amplitudes, with $R$ equal to the reference range. Amplitude SSIM instead uses the same fields after one global mean-amplitude match of the reconstruction to the reference. No support or background mask is applied.

For the phase metrics, two pixels are trimmed from each edge of each held-out LCLS XPP array. Let $\phi_{\mathrm{rec},j}=\arg U_{\mathrm{rec},j}$ and $\phi_{\mathrm{ref},j}=\arg U_{\mathrm{ref},j}$ denote the principal phases. Unweighted least-squares planes are fitted to and removed from the reconstructed and reference phases separately:
\begin{equation}
\begin{aligned}
(a_s,b_s,c_s)
&=
\operatorname*{arg\,min}_{a,b,c}
\sum_{j=1}^{N_\Omega}
\left[\phi_{s,j}-(a x_j+b y_j+c)\right]^2,
\qquad s\in\{\mathrm{rec},\mathrm{ref}\},\\
\widetilde{\phi}_{s,j}
&=
\phi_{s,j}-(a_s x_j+b_s y_j+c_s),\\
\Delta\phi_j
  &=\arg\!\left\{
  \exp\!\left[\mathrm{i}\left(
  \widetilde{\phi}_{\mathrm{rec},j}
  -\widetilde{\phi}_{\mathrm{ref},j}\right)\right]
  \right\},\\
\mathrm{RMS}_{\phi}
  &=
  \sqrt{\frac{1}{N_\Omega}\sum_{j=1}^{N_\Omega}
  \left(\Delta\phi_j\right)^2}.
\end{aligned}
\end{equation}
The residual is wrapped to $(-\pi,\pi]$ before the unweighted RMS is evaluated. Reference intensity is not used as a weight, and no binary support mask or fine registration is applied. The resulting quantity measures agreement with the reference phase after plane alignment, rather than absolute phase accuracy.

\begin{figure}[htbp]
  \centering
  \includegraphics[width=0.92\linewidth]{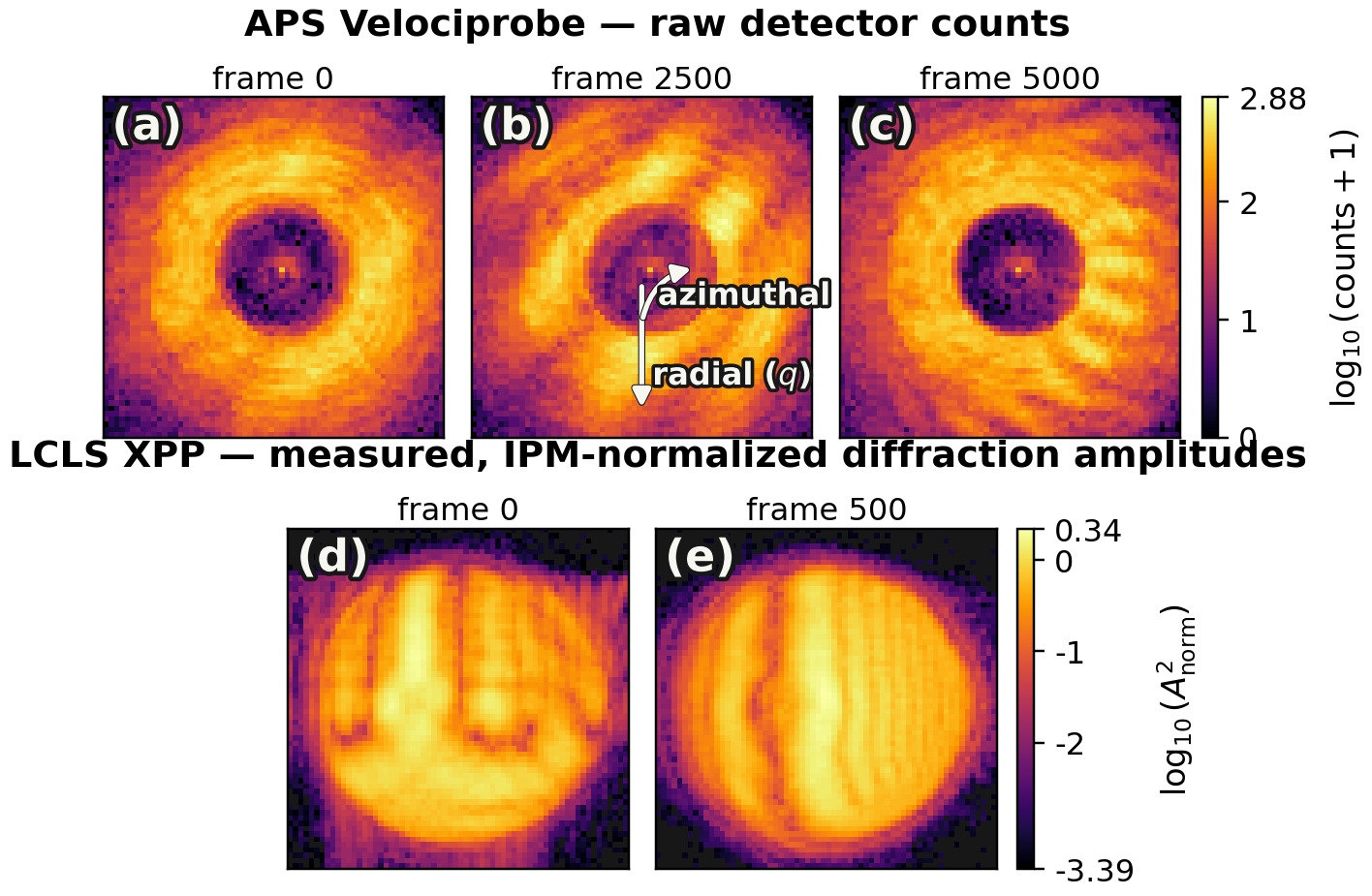}
  \caption{Experimental diffraction data. (a)--(c) APS Velociprobe (fly64) raw-count frames 0, 2500, and 5000 displayed as $\log_{10}(\mathrm{counts}+1)$. (d),(e) LCLS XPP measured, IPM-normalized diffraction-amplitude frames 0 and 500 displayed as $\log_{10}(A_{\mathrm{norm}}^2)$ with exact-zero pixels masked.}
  \label{fig:raw_diffraction}
\end{figure}

Figure~\ref{fig:raw_diffraction} shows representative inputs: APS raw detector counts and LCLS XPP IPM-normalized diffraction amplitudes.

\section{Results}

We report results on the APS Velociprobe Siemens-star data, the LCLS XPP dataset, the synthetic Siemens-star dataset, and the synthetic line-pattern dataset; see Methods for dataset definitions and evaluation protocol.

\subsection{Reconstruction Quality}

\begin{figure}[p]
    \centering
    \includegraphics[width=\linewidth]{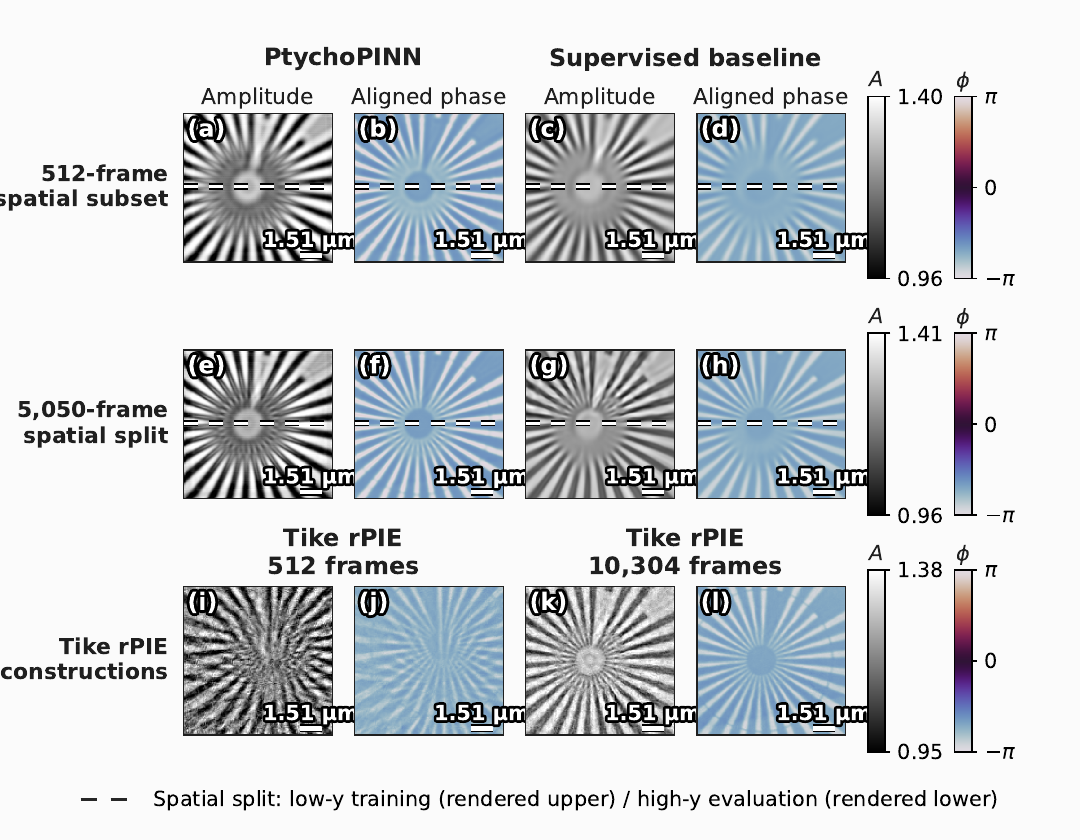}
    \caption{Full-scan APS Siemens-star reconstructions trained on 512 and 5,050 frames, compared with Tike (rPIE) object references reconstructed from 512 and 10,304 frames. All learned models are evaluated at 10,304 positions. The dashed line marks the spatial training/evaluation boundary; both models train on frames from the training half only, the 512-frame model on a random subsample of it. Rows share amplitude scales and phase spans $[-\pi,\pi]$.}
    \label{fig:smalldat}
\end{figure}

Figure~\ref{fig:smalldat} compares reconstructions on the APS Siemens-star data using a spatial holdout: one half of the scan is used for training and the other half for testing, with models trained on either all 5,050 training-half frames or a random 512-frame subsample, then evaluated over the full scan. At 5,050 frames, the supervised baseline reconstructs the training region well but degrades on held-out positions, whereas PtychoPINN maintains consistent quality across both. At 512 frames, this train--test gap widens further for the supervised baseline. On the synthetic Siemens-star dataset (simulated from APS Siemens-star reconstructions), PtychoPINN also attains higher phase fidelity than the supervised baseline; see Table~\ref{tab:results_16384}.

Against known synthetic Siemens-star ground truth, PtychoPINN at 16,384 training patterns reaches amplitude and phase SSIM values of 0.955 and 0.962, compared with 0.930 and 0.912 for the supervised baseline (Table~\ref{tab:results_16384}). 

\begin{table}[htbp]
\centering
\caption{Synthetic Siemens-star reconstruction quality (PSNR and SSIM) at 16,384 training patterns against simulated-object ground truth. Values are mean $\pm$ standard deviation over five runs; best values are highlighted in \textcolor{bestval}{green}.}
\label{tab:results_16384}
\begin{adjustbox}{max width=\textwidth}
\begin{tabular}{@{}llcccc@{}}
\toprule
& & \multicolumn{2}{c}{PSNR (dB)} & \multicolumn{2}{c}{SSIM} \\
\cmidrule(lr){3-4} \cmidrule(lr){5-6}
Dataset & Method & Amplitude & Phase & Amplitude & Phase \\
\midrule
\multirow{2}{*}{Synthetic Siemens star}
& Supervised baseline
& $84.83 \pm 0.23$
& $68.62 \pm 0.02$
& $0.930 \pm 0.002$
& $0.912 \pm 0.003$ \\
& PtychoPINN
& \textcolor{bestval}{$\mathbf{85.53 \pm 0.02}$}
& \textcolor{bestval}{$\mathbf{70.54 \pm 0.06}$}
& \textcolor{bestval}{$\mathbf{0.955 \pm 0.001}$}
& \textcolor{bestval}{$\mathbf{0.962 \pm 0.001}$} \\
\bottomrule
\end{tabular}
\end{adjustbox}
\end{table}

\subsection{Overlap and Probe-Structure Comparison}

\begin{figure}[htbp]
    \centering
    \includegraphics[width=\textwidth]{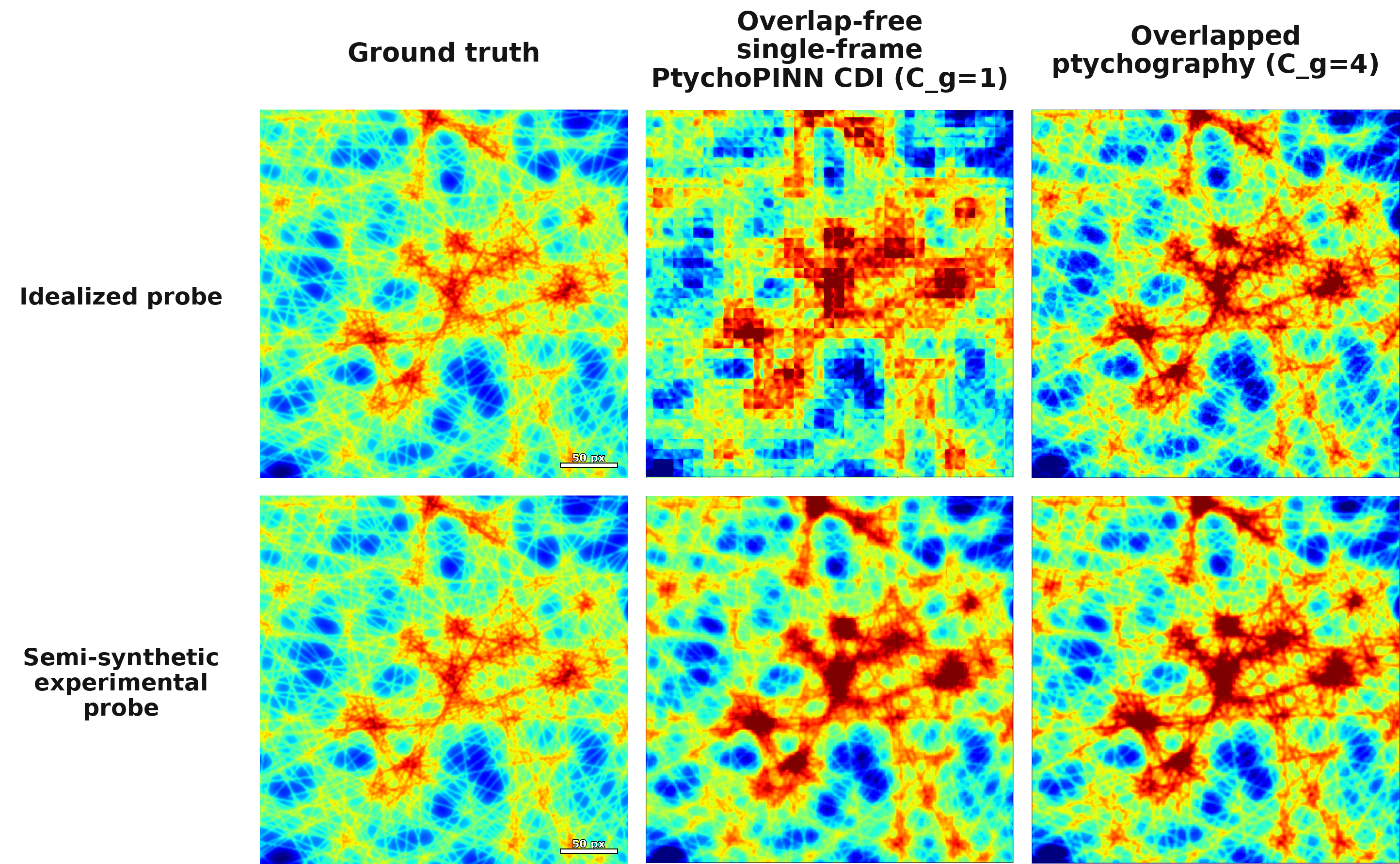}
    \caption{Synthetic line-pattern reconstructions across probe structure and reconstruction group size. The idealized probe is a smoothed disk with uniform phase; the experimental-probe condition applies the curved probe reconstructed from the off-focus LCLS XPP geometry. The $C_g=1$ column uses no cross-frame overlap constraint. Coordinates are dimensionless.}
    \label{fig:recon_2x2}
\end{figure}

In overlap-free operation, we set the group size parameter to a single diffraction frame ($C_g = 1$). Reconstruction then relies entirely on the diffraction likelihood and the known probe structure (a curved wavefront from the off-focus sample position). Figure~\ref{fig:recon_2x2} illustrates the synthetic object ground truth alongside overlap-free single-frame reconstructions and overlapped ptychographic reconstructions ($C_g=4$), all generated by PtychoPINN. Each overlap condition is run on two simulated datasets, one with an idealized flat-phase probe and the other with an experimentally derived curved probe. Table~\ref{tab:sim_lines_metrics} reports metrics for the four combinations of overlap ($C_g=1$ vs.\ $C_g=4$) and probe structure (flat vs.\ curved). With the curved probe, removing overlap reduces amplitude SSIM by 0.064 (0.968 to 0.904) and PSNR by 4.14 dB (73.03 to 68.89). With the idealized flat-phase probe, the degradation is much larger: SSIM drops by 0.332 (0.952 to 0.620) and PSNR by 10.67 dB (71.34 to 60.67). Figure~\ref{fig:recon_2x2} shows the same pattern qualitatively: overlap-free reconstruction recovers the line-pattern object only when the curved probe is used, whereas the idealized flat-phase probe fails to anchor the reconstruction without overlap.

\begin{table}[htbp]
\centering
\caption{Synthetic line-pattern amplitude reconstruction metrics on the test split. Ground truth is the simulated object (amplitude only; the object has constant zero phase).}
\label{tab:sim_lines_metrics}
\begin{tabular}{@{}lcc@{}}
\toprule
Case & PSNR (dB) & SSIM \\
\midrule
overlap-free ($C_g=1$) + idealized probe & 60.67 & 0.620 \\
overlap-free ($C_g=1$) + experimental probe & 68.89 & 0.904 \\
overlap ($C_g=4$) + idealized probe & 71.34 & 0.952 \\
overlap ($C_g=4$) + experimental probe & 73.03 & 0.968 \\
\bottomrule
\end{tabular}
\end{table}

\subsection{Photon-Limited Performance}

\begin{figure}[htbp]
    \centering
    \begin{subfigure}[t]{0.28\linewidth}
        \centering
        \textbf{$10^9$ photons/frame}\\[2pt]
        \includegraphics[width=0.62\linewidth]{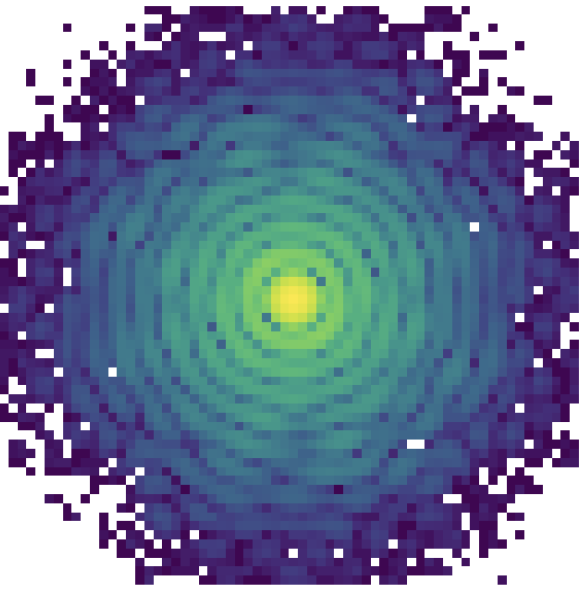}\\[6pt]
        \textbf{$10^4$ photons/frame}\\[2pt]
        \includegraphics[width=0.62\linewidth]{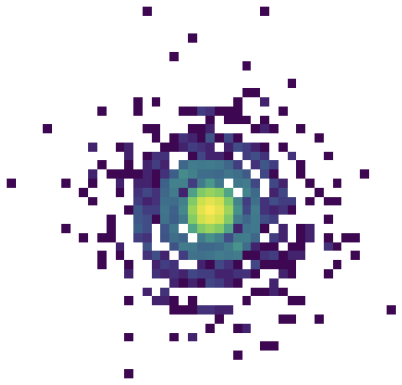}
        \caption{Simulated diffraction patterns.}
        \label{fig:dose_diffraction}
    \end{subfigure}\hfill
    \begin{subfigure}[t]{0.68\linewidth}
        \centering
        \begin{tabular}{@{}c@{\hspace{6pt}}c@{}}
            \textbf{Amplitude MAE} & \textbf{Poisson NLL} \\[3pt]
            \multicolumn{2}{c}{\textbf{$10^9$ photons/frame}} \\
            \includegraphics[width=0.42\linewidth]{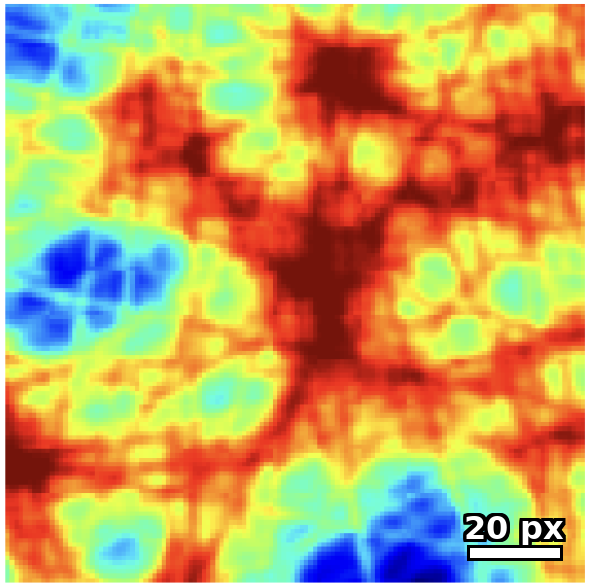} &
            \includegraphics[width=0.42\linewidth]{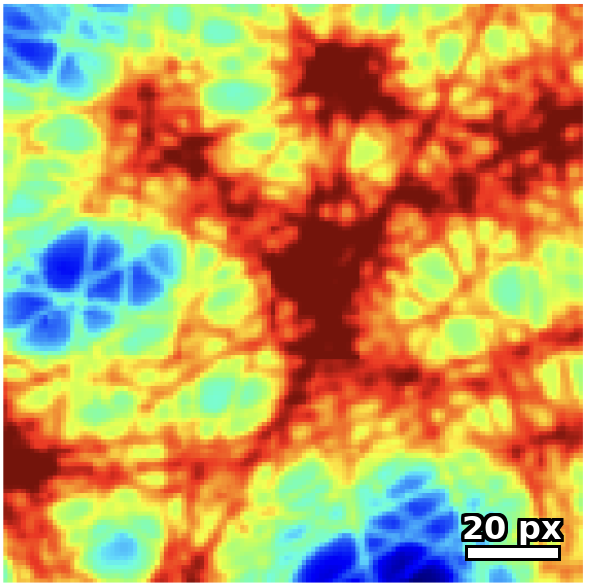} \\[6pt]
            \multicolumn{2}{c}{\textbf{$10^4$ photons/frame}} \\
            \includegraphics[width=0.42\linewidth]{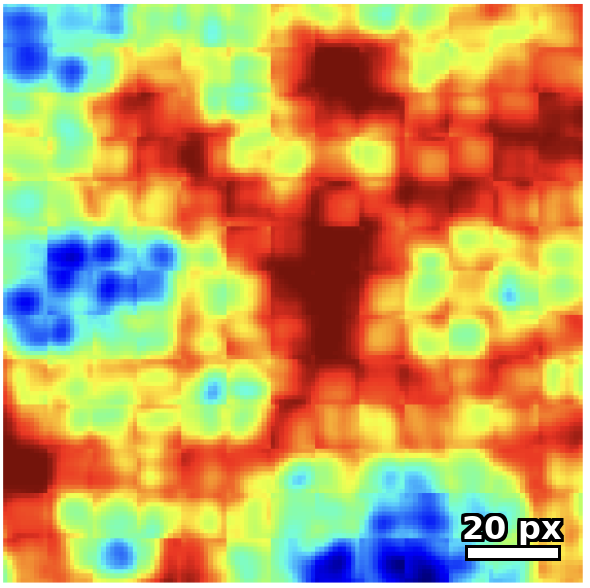} &
            \includegraphics[width=0.42\linewidth]{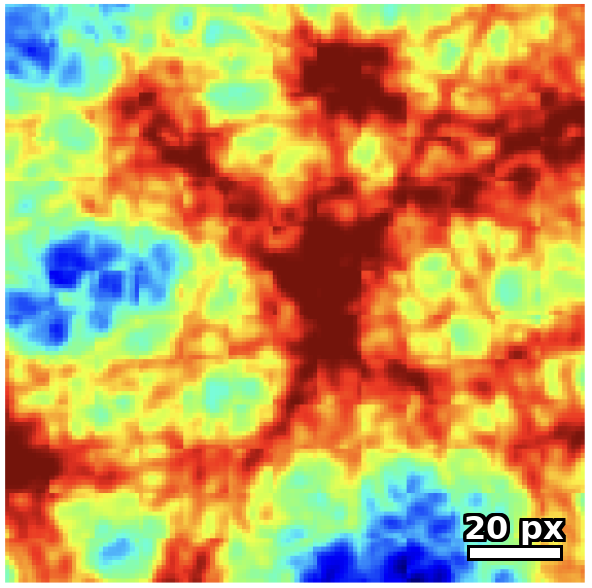}
        \end{tabular}
        \caption{Reconstructions by objective.}
        \label{fig:dose_recon}
    \end{subfigure}

    \medskip
    \begin{subfigure}[t]{0.62\linewidth}
        \centering
        \includegraphics[width=\linewidth]{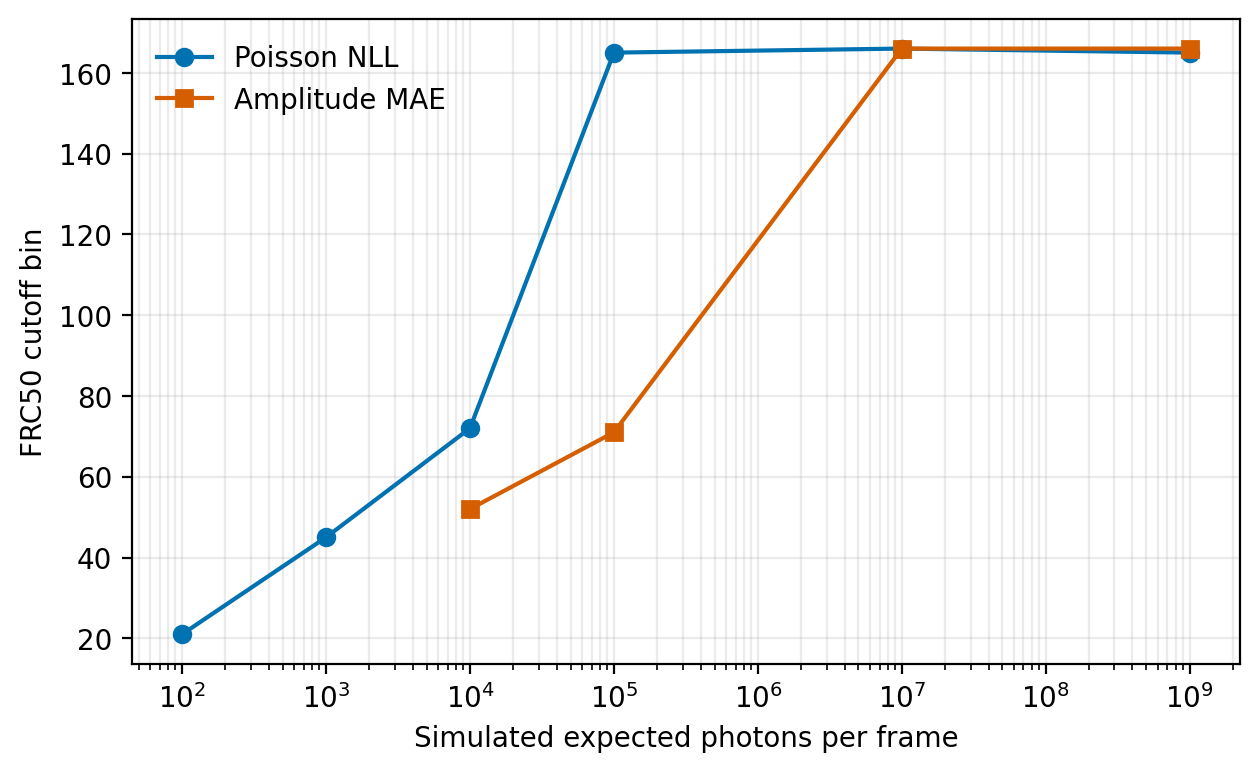}
        \caption{FRC50 cutoff versus simulated expected photons.}
        \label{fig:dose_frc}
    \end{subfigure}
    \caption{Photon-limited reconstruction on the synthetic line-pattern dataset. FRC50 is the first crossing below 0.5 and is reported as a spatial-frequency cutoff bin versus simulated expected photons per diffraction frame.}
    \label{fig:dose}
\end{figure}

Figure~\ref{fig:dose} compares resolution using the 50\% Fourier ring correlation criterion (FRC50) as a function of photon dose for Poisson NLL versus MAE training objectives. At low dose (${\sim}10^4$ photons/frame), the Poisson NLL matches the resolution that MAE reaches only at roughly $10\times$ higher dose, corresponding to an order-of-magnitude improvement in dose efficiency. This advantage arises because the Poisson likelihood correctly models photon-counting noise, preserving sensitivity to the low-count, high-$q$ components that carry fine spatial detail but are overwhelmed by bright-pixel residuals under an amplitude MAE.

\subsection{Data Efficiency}

\begin{figure}[htbp]
    \centering
    \includegraphics[width=0.8\textwidth]{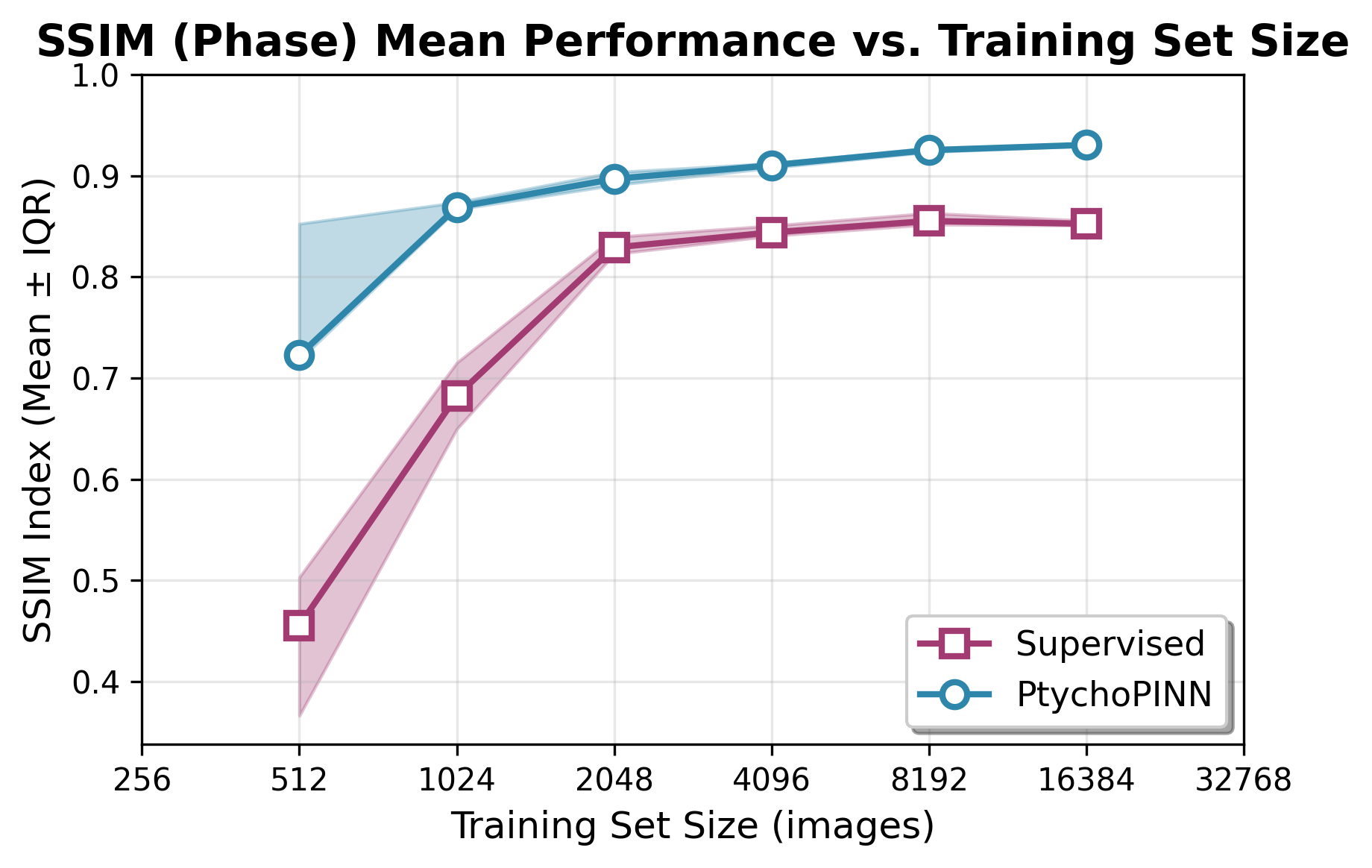}
    \caption{Phase structural similarity of PtychoPINN and the supervised baseline as a function of training-set size.}
    \label{fig:ssim}
\end{figure}

Figure~\ref{fig:ssim} shows reconstruction quality (phase SSIM) as a function of training-set size. PtychoPINN maintains high fidelity (SSIM $>0.85$) down to 1,024 diffraction patterns, whereas the supervised baseline degrades rapidly below 2,048 training samples. At small training-set sizes, PtychoPINN matches the supervised baseline with roughly an order of magnitude less training data, consistent with the physics constraints enforced by the forward model acting as an effective prior for this inverse problem.

\subsection{Cross-Facility Evaluation}
\begin{figure}[t]
  \centering
  \includegraphics[width=\linewidth]{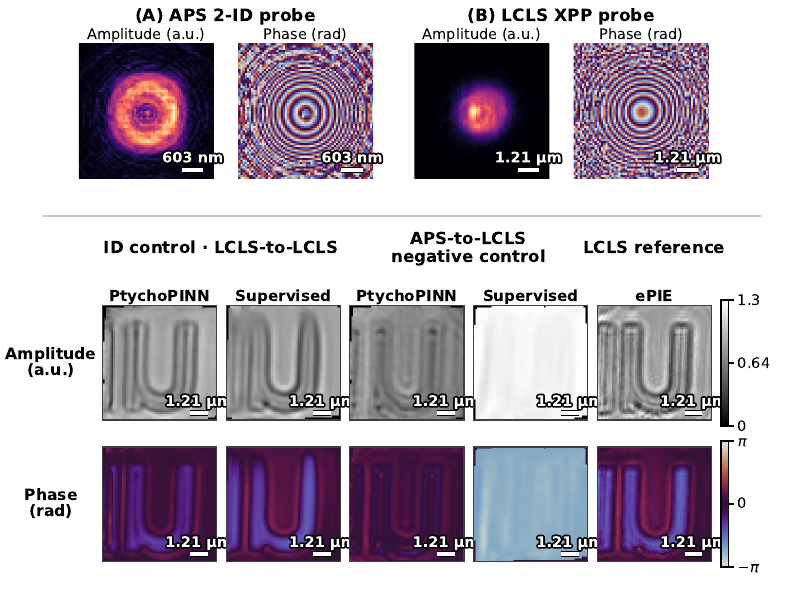}
  \caption{Aligned-field comparison of PtychoPINN and the distinct supervised baseline for in-distribution LCLS XPP reconstruction (train B $\rightarrow$ test B) and APS-to-LCLS evaluation (train A $\rightarrow$ test B), with the same-acquisition ePIE LCLS XPP reference. Reconstruction and reference fields share spatial extents and modality-specific display scales; phase spans $[-\pi,\pi]$~rad.}
  \label{fig:ood_comparison}
\end{figure}

\begin{figure}[p]
  \centering
  \includegraphics[width=0.78\linewidth]{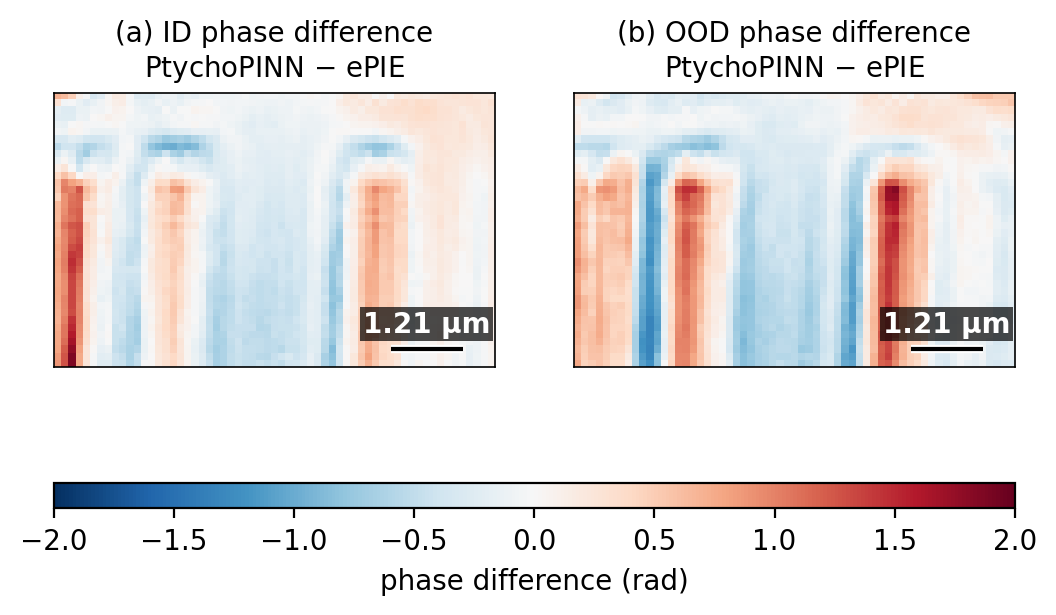}
  \caption{Plane-aligned PtychoPINN-minus-ePIE phase differences on the held-out LCLS XPP test half for in-distribution LCLS training and APS-to-LCLS application. Two edge pixels are trimmed, an unweighted plane is removed separately from each principal-phase field, and the residual is wrapped to $(-\pi,\pi]$. No support mask or fine registration is applied.}
  \label{fig:phase_analysis}
\end{figure}

Figure~\ref{fig:ood_comparison} compares an in-distribution reconstruction of LCLS data (train LCLS XPP, test LCLS XPP) with an out-of-distribution transfer (train APS, test LCLS XPP). Under APS$\!\to\!$LCLS shift, the supervised baseline largely collapses, whereas PtychoPINN preserves recognizable edge structure, albeit with poor fidelity. The reference column is the ePIE reconstruction of the LCLS data. Table~\ref{tab:fig5_ood_metrics} reports the held-out metrics for both conditions, and Fig.~\ref{fig:phase_analysis} shows the corresponding plane-aligned phase residuals.

Complex Fourier-domain agreement with ePIE over the held-out LCLS XPP region depends strongly on crop-window and radial-smoothing choices (Supplementary Fig.~S2): the first nominal 0.5 crossing for the in-distribution field ranges from 0.62 to 0.96 of Nyquist across processing choices, so these curves serve as frequency-resolved agreement diagnostics rather than resolution estimates.



\begin{table}[p]
  \centering
  \caption{Reconstruction metrics on the held-out LCLS XPP test-half region against the ePIE reference after trimming two pixels from each edge. Amplitude MSE and PSNR use unscaled, unmasked fields; amplitude SSIM uses global mean matching. Phase RMS uses a wrapped residual between separately plane-aligned fields. No support mask or fine registration is applied. The ePIE reference also supplied the supervised baseline's target patches. Bold marks the better model within each condition.}
  \label{tab:fig5_ood_metrics}
\begin{adjustbox}{max width=\columnwidth}
\begin{tabular}{llrrrr}
\toprule
Condition & Model & Amp. MSE & Amp. PSNR & Amp. SSIM & Phase RMS (rad) \\
\midrule
ID & PtychoPINN & 0.01501 & 17.539 & 0.4955 & 0.39409 \\
ID & Supervised baseline & \textbf{0.01305} & \textbf{18.147} & \textbf{0.538} & \textbf{0.27266} \\
OOD & PtychoPINN & \textbf{0.03224} & \textbf{14.219} & \textbf{0.1757} & \textbf{0.54317} \\
OOD & Supervised baseline & 0.2476 & 5.366 & 0.07236 & 0.63093 \\
\bottomrule
\end{tabular}
\end{adjustbox}
\end{table}

\subsection{Computational Performance}

On an NVIDIA RTX 3090, feedforward inference processes approximately $6.1\times10^3$ frames/s at $64\times64$ and $2.6\times10^3$ frames/s at $128\times128$. For the evaluated 10,304-frame workload, the $128\times128$ throughput corresponds to approximately 4.0~s for PtychoPINN reconstruction. The measured PtyChi LSQ-ML time of 1.444~s per epoch corresponds to 144.4~s for 100 epochs, giving PtychoPINN an approximately $36\times$ end-to-end reconstruction-time advantage for this workload~\cite{Du2025PtyChi,Odstrcil2018LSQML}.

\section{Discussion}

\subsection*{Overlap as an optional reconstruction constraint}
With a curved-wavefront probe, PtychoPINN reconstructs each illuminated region of an extended sample from a single diffraction pattern without using neighboring measurements to constrain phase retrieval. The synthetic line-pattern ablation (Fig.~\ref{fig:recon_2x2}, Table~\ref{tab:sim_lines_metrics}) demonstrates that overlap improves reconstruction fidelity but is not required for recovery. This reflects the well-known symmetry-breaking property of probe phase diversity that makes Fresnel CDI possible. 

In this context, overlap~\cite{Bunk2008Overlap,Edo2013PRA} and probe phase diversity~\cite{Williams2006FresnelCDI,Levitan2020Randomized} act as partially substitutable sources of constraint. This separates two roles that conventional ptychography couples: overlap as a source of phase-retrieval redundancy and scan density as a choice governing spatial coverage and dose. In practice, acquisitions can use fewer positions, less overlap, or---with a curved-wavefront probe providing phase diversity---no scanning at all, reducing acquisition time and total dose. These properties are particularly relevant for dynamic or radiation-sensitive samples at high-repetition-rate sources, where overlap requirements and photon budget are simultaneous constraints.

Self-supervision removes the need for object ground truth but not the need to specify the illumination. In single-frame recovery, the fixed probe anchors the factorization of the exit wave into illumination and object, so any probe error is absorbed into the reconstructed object. Joint probe prediction would be subject to blind object--probe ambiguity~\cite{Williams2006FresnelCDI,Abbey2008KeyholeCDI,gao2023iterative,Marchesini2003Shrinkwrap}, but amortized inference of the probe over the course of an entire training run is a plausible direction for improvement.


\subsection*{Dose efficiency}
Single-frame reconstruction and Poisson-likelihood training reduce dose---the fundamental limit on resolution for radiation-sensitive specimens~\cite{Howells2009Dose}---through distinct mechanisms. Removing the overlap requirement allows a fixed field of view to be covered with fewer exposures, lowering total acquisition dose at fixed exposure per frame. The Poisson objective separately improves photon use within each exposure: in count-resolved simulation, it matches the FRC50 resolution of the detector-amplitude MAE objective with one-tenth the photon dose (Fig.~\ref{fig:dose_frc}), a tenfold improvement in simulated photon-dose efficiency. (This loss-function result is established in simulation; the normalized APS and LCLS diffraction amplitudes do not provide an experimental photon-dose calibration.)

\subsection*{Diffraction-space supervision}
The forward model provides a dense physical constraint per measurement: each diffraction pattern constrains the full exit-wave amplitude, and the Poisson NLL weights every detector pixel, including the low-count, high-$q$ pixels where fine spatial detail resides (Fig.~\ref{fig:dose}). Real-space supervision instead fits the network to a single reference reconstruction that already carries ambiguities intrinsic to the inverse problem, such as global phase offsets. Because such nuisance parameters are not uniquely determined by the data, a supervised network can overfit to them; this mechanism is consistent with both the supervised baseline's train--test gap on held-out scan positions (Fig.~\ref{fig:smalldat}) and its loss of recognizable object structure under APS-to-LCLS application (Fig.~\ref{fig:ood_comparison}, Table~\ref{tab:fig5_ood_metrics}). The same mechanism also plausibly underlies PtychoPINN's observed data efficiency (Fig.~\ref{fig:ssim}).

\subsection*{Experimental validation and cross-facility behavior}

The APS spatial-holdout comparison (Fig.~\ref{fig:smalldat}) suggests that PtychoPINN learns a transferable inverse mapping rather than memorizing the training region: unlike the supervised baseline, it reconstructs held-out scan positions as well as training positions, even when the training set is reduced from 5,050 to 512 frames. In the spatial holdout, however, only the scan positions differ between training and evaluation; the sample and the forward model are the same. APS-to-LCLS application changes the sample morphology, effective probe, object-plane sampling, and measurement statistics all at once, and the network receives none of these as inputs. Under this compound shift, PtychoPINN remains closer to the LCLS ePIE reference than the supervised baseline and retains the test-pattern geometry in both amplitude and phase (Fig.~\ref{fig:ood_comparison}, Table~\ref{tab:fig5_ood_metrics}), but amplitude SSIM remains low, plane-aligned phase RMS increases relative to the in-distribution control, and the phase residuals strengthen along the test-pattern boundaries (Fig.~\ref{fig:phase_analysis}). These results demonstrate partial retention of object structure under a compound distribution shift, not invariance to any individual factor. Establishing broader generalization will require controlled tests that vary morphology and acquisition conditions independently.

\subsection*{Computational implications and open problems}
Once trained, PtychoPINN reduces each subsequent reconstruction to feedforward inference, whereas LSQ-ML performs a new iterative optimization for each dataset. The measured $36\times$ end-to-end speedup therefore reflects a reduction in per-dataset reconstruction cost.

The experimental reconstructions reported here use $N=64$ model grids, a fixed pre-estimated probe, and supplied scan positions. Holding the probe and positions fixed isolates single-frame object recovery from calibration refinement, but it also makes probe drift and geometry errors sources of forward-model mismatch. Joint refinement of the probe and scan positions, following established iterative practice~\cite{Thibault2009Probe,GuizarSicairos2008TTD,Maiden2012Annealing,Zhang2013Position,Du2024ProbePosition}, is therefore the direct route to improving robustness across experiments. Scaling to larger reconstruction grids is a distinct architectural problem: the convolutional inverse backbone is the dominant scaling bottleneck, motivating backbones that combine global mixing of diffraction features with local real-space synthesis. Fourier neural operator (FNO) backbones~\cite{Li2021FNO} are a natural candidate, because global spectral mixing is expected to scale more favorably with grid size $N$ than convolutional receptive fields. The modular separation of inverse backbone, differentiable forward model, and loss should also permit further speedups from mixed-precision and architecture-level optimization without changing the training procedure, and should simplify adaptation to other coherent imaging modalities.


\section{Conclusions}
We extend PtychoPINN from overlapped ptychography to single-frame Fresnel CDI of extended samples. Each illuminated region is recovered from its own diffraction pattern, removing neighboring-frame overlap as a reconstruction requirement and permitting sparser scans. When overlap is available, the same framework uses it as an additional constraint, unifying the two acquisition regimes within one physics-constrained, self-supervised formulation.

In the synthetic line-pattern study, single-frame reconstructions with the curved probe retained amplitude SSIM above 0.90, increasing to 0.952--0.968 when overlap was used. This single-frame capability permits a fixed field of view to be sampled with fewer exposures, and hence at lower total dose for a fixed exposure per frame. A Poisson negative log likelihood objective provides a separate gain, matching the FRC50 cutoff of amplitude MAE with tenfold fewer simulated photons per frame. Once trained, PtychoPINN reconstructed the evaluated 10,304-frame, $128\times128$ workload in approximately 4.0~s, a $36\times$ end-to-end speedup over 100 epochs of PtyChi LSQ-ML.

Reconstructions from APS and LCLS measurements demonstrate that the single-frame approach extends to experimental data. When the APS-trained model was applied to LCLS without retraining, PtychoPINN agreed more closely with the LCLS ePIE reference than the supervised baseline, although substantial amplitude and low-order phase mismatch remained. Joint refinement of the probe and scan positions is a plausible next step toward a single-frame reconstruction pipeline that retains its accuracy as experimental conditions change. 

\section*{Appendix A: Configuration Parameters}

\begin{table}[!htbp]
\centering
\caption{Model, training, and evaluation settings.}
\label{tab:config_params}
\small
\renewcommand{\arraystretch}{1.08}
\begin{tabularx}{\textwidth}{@{}p{0.23\textwidth}X@{}}
\toprule
\textbf{Setting} & \textbf{Value} \\
\midrule
Detector/model grid & $N=64$ for reconstruction-quality experiments; $N=128$ for the throughput measurement. \\
Single-frame and overlap grouping & $C_g=1$ for experimental reconstructions; the synthetic overlap ablation compares $C_g=1$ with adjacent four-frame groups ($C_g=4$). \\
PtychoPINN objectives & Detector-amplitude MAE and detector-intensity Poisson NLL. APS/LCLS models use $(w_{\mathrm{MAE}},w_{\mathrm{Poiss}})=(0,1)$, $N_{\mathrm{eff}}=10^9$, $s_0=988.2117$ at $N=64$, and trainable $s_\alpha$; the scale is imposed rather than measured photon count. The overlap ablation uses MAE; the synthetic photon-count sweep compares both objectives. \\
Supervised baseline & Distinct PtychoNN-style dual-head U-Net; one diffraction input, no coordinates, direct amplitude/phase MAE. \\
Optimization & Adam, initial rate $10^{-3}$; reduce-on-plateau factor $0.5$, patience $2$, floor $10^{-4}$. \\
Training schedule & Epoch caps $50$ or $60$ by study; batch size $16$. \\
Splits & $5\%$ of selected training examples for Keras validation; separate dataset-specific held-out evaluation sets. \\
Probe and positions & Fixed probe and supplied scan positions during object-network training; neither is jointly refined. \\
Decoder outputs & Central sigmoid amplitude and $\pi\tanh$ phase; four-channel swish peripheral continuation with nearest-neighbor upsampling. \\
Architecture and grouping defaults & Network width multiplier \texttt{n\_filters\_scale} $=2$; encoder depth $d=3$--$5$ (resolution-dependent); nearest-neighbor count $K=4$; oversampling restriction \texttt{pad\_object} enabled ($N/2\times N/2$ central tile); no probe mask; probe Gaussian smoothing $\sigma=0$; scan-step offset 4\,px in synthetic grid scans. \\
Hardware and software & NVIDIA RTX 3090 (24\,GB), TensorFlow/Python 3.10. \\
\bottomrule
\end{tabularx}
\end{table}

\begin{table}[!htbp]
\centering
\caption{Symbol definitions.}
\label{tab:symbols}
\small
\renewcommand{\arraystretch}{1.08}
\begin{tabularx}{\textwidth}{@{}lp{0.30\textwidth}X@{}}
\toprule
\textbf{Symbol} & \textbf{Type / structure} & \textbf{Description} \\
\midrule
$a'_k$ & $N\times N$ nonnegative real array & Measured detector amplitude for frame $k$ \\
$x_k$ & $N\times N$ real array & Normalized diffraction amplitude, Eq.~\eqref{eq:norm} \\
$a_k$, $n_k$ & $N\times N$ real arrays & External amplitude and intensity targets, Eq.~\eqref{eq:internalscale} \\
$\vec r_k$ & 2D position vector & Supplied absolute scan position of frame $k$ \\
$\bar{\vec r}$ & 2D position vector & Centroid of a reconstruction group \\
$\vec r^{\,\mathrm{rel}}_k$ & 2D offset vector & Stored model offset $\bar{\vec r}-\vec r_k$, Eq.~\eqref{eq:relativecoords} \\
$s_\alpha=e^{\alpha_{\log}}$ & Scalar (trainable or fixed) & Shared internal amplitude scale \\
$s_0$ & Scalar (fixed) & Dataset-to-target amplitude factor, Eq.~\eqref{eq:alphaloginit} \\
$N_{\mathrm{eff}}$ & Scalar & Configured mean integrated target intensity per frame \\
$P(\vec r)$ & $N\times N$ complex array & Fixed effective probe \\
$O_k$ & $N\times N$ complex array & Object patch decoded by the network $G$ \\
$O_{\mathrm{region}}$ & $M\times M$ complex array & Merged object representation, Eq.~\eqref{eq:constraintmap} \\
$O'_k$ & $N\times N$ complex array & Patch extracted from $O_{\mathrm{region}}$ in the forward map \\
$\Psi_k$ & $N\times N$ complex array & Unnormalized discrete Fourier transform of the exit wave \\
$A^{(0)}_k$ & $N\times N$ real array & Internally normalized detector amplitude \\
$\hat A_k$ & $N\times N$ real array & Predicted target-scale detector amplitude \\
$\mu_k$ & $N\times N$ real array & Predicted Poisson intensity rate \\
\bottomrule
\multicolumn{3}{@{}l}{\footnotesize $N$: detector/patch side length; $C_g$: joint reconstruction group size; $M$: merged-canvas side length.} \\
\end{tabularx}
\end{table}

\clearpage
\begin{backmatter}
\bmsection{Funding}
This work was supported by the U.S. Department of Energy, Laboratory Directed Research and Development program at SLAC National Accelerator Laboratory, under Contract No. DE-AC02-76SF00515.

It was also supported by the U.S. Department of Energy (DOE), Office of Science, Basic Energy Sciences, through the Collaborative Machine Learning Platform for Scientific Discovery 2.0 award. Use of the Advanced Photon Source was supported by the U.S. Department of Energy, Office of Science, Office of Basic Energy Sciences, under Contract No. DE-AC02-06CH11357.

\bmsection{Disclosures}
The authors declare no conflicts of interest.

\bmsection{Data Availability}
The LCLS XPP dataset and PtychoPINN model and training code are publicly available at \url{https://github.com/hoidn/PtychoPINN}. The APS fly64 dataset, trained model weights, derived metric arrays, and analysis manifests are held by the authors and are available upon reasonable request, subject to facility data policies.
\end{backmatter}

\bibliography{references}

\begin{thebibliography}{10}
\newcommand{\enquote}[1]{``#1''}

\bibitem{LCLSIIHE_DesignPerf}
{SLAC National Accelerator Laboratory}, \enquote{{LCLS-II-HE Design and
  Performance},}
  \url{https://lcls.slac.stanford.edu/lcls-ii-he/design-and-performance}
  (2023). Accessed: 2026-09-09.

\bibitem{GuizarSicairos2021PhysicsToday}
M.~Guizar-Sicairos and P.~Thibault, \enquote{Ptychography: A solution to the
  phase problem,} {\protect\JournalTitle{Physics Today}} \textbf{74}, 42--48
  (2021).

\bibitem{Pfeiffer2018XrayPtycho}
F.~Pfeiffer, \enquote{{X}-ray ptychography,} {\protect\JournalTitle{Nature
  Photonics}} \textbf{12}, 9--17 (2018).

\bibitem{Nugent2010AdvPhys}
K.~A. Nugent, \enquote{Coherent methods in the {X}-ray sciences,}
  {\protect\JournalTitle{Advances in Physics}} \textbf{59}, 1--99 (2010).

\bibitem{Chapman2010NatPhoton}
H.~N. Chapman and K.~A. Nugent, \enquote{Coherent lensless {X}-ray imaging,}
  {\protect\JournalTitle{Nature Photonics}} \textbf{4}, 833--839 (2010).

\bibitem{Miao2015Science}
J.~Miao, T.~Ishikawa, I.~K. Robinson, and M.~M. Murnane, \enquote{Beyond
  crystallography: diffractive imaging using coherent x-ray light sources,}
  {\protect\JournalTitle{Science}} \textbf{348}, 530--535 (2015).

\bibitem{Giewekemeyer2010PNAS}
K.~Giewekemeyer, P.~Thibault, S.~Kalbfleisch \emph{et~al.},
  \enquote{Quantitative biological imaging by ptychographic x-ray diffraction
  microscopy,} {\protect\JournalTitle{Proceedings of the National Academy of
  Sciences}} \textbf{107}, 529--534 (2010).

\bibitem{Dierolf2010Nature}
M.~Dierolf, A.~Menzel, P.~Thibault \emph{et~al.}, \enquote{Ptychographic
  {X}-ray computed tomography at the nanoscale,}
  {\protect\JournalTitle{Nature}} \textbf{467}, 436--439 (2010).

\bibitem{Holler2017Nature}
M.~Holler, M.~Guizar-Sicairos, E.~H.~R. Tsai \emph{et~al.},
  \enquote{High-resolution non-destructive three-dimensional imaging of
  integrated circuits,} {\protect\JournalTitle{Nature}} \textbf{543}, 402--406
  (2017).

\bibitem{Jiang2018Nature}
Y.~Jiang, Z.~Chen, Y.~Han \emph{et~al.}, \enquote{Electron ptychography of {2D}
  materials to deep sub-{\aa}ngstr{\"o}m resolution,}
  {\protect\JournalTitle{Nature}} \textbf{559}, 343--349 (2018).

\bibitem{Williams2006FresnelCDI}
G.~J. Williams, H.~M. Quiney, B.~B. Dhal, \emph{et~al.}, \enquote{Fresnel
  coherent diffractive imaging,} {\protect\JournalTitle{Physical Review
  Letters}} \textbf{97}, 025506 (2006).

\bibitem{Abbey2008KeyholeCDI}
B.~Abbey, K.~A. Nugent, G.~J. Williams, \emph{et~al.}, \enquote{Keyhole
  coherent diffractive imaging,} {\protect\JournalTitle{Nature Physics}}
  \textbf{4}, 394--398 (2008).

\bibitem{Vine2009PFCDI}
D.~J. Vine, G.~J. Williams, B.~Abbey, \emph{et~al.}, \enquote{Ptychographic
  fresnel coherent diffractive imaging,} {\protect\JournalTitle{Physical Review
  A}} \textbf{80}, 063823 (2009).

\bibitem{Levitan2020Randomized}
A.~L. Levitan, K.~Keskinbora, U.~T. Sanli \emph{et~al.}, \enquote{Single-frame
  far-field diffractive imaging with randomized illumination,}
  {\protect\JournalTitle{Optics Express}} \textbf{28}, 37103--37117 (2020).

\bibitem{GerchbergSaxton1972}
R.~W. Gerchberg and W.~O. Saxton, \enquote{A practical algorithm for the
  determination of phase from image and diffraction plane pictures,}
  {\protect\JournalTitle{Optik}} \textbf{35}, 237--246 (1972).

\bibitem{Fienup1982}
J.~R. Fienup, \enquote{Phase retrieval algorithms: a comparison,}
  {\protect\JournalTitle{Applied Optics}} \textbf{21}, 2758--2769 (1982).

\bibitem{Marchesini2007Review}
S.~Marchesini, \enquote{Invited article: A unified evaluation of iterative
  projection algorithms for phase retrieval,} {\protect\JournalTitle{Review of
  Scientific Instruments}} \textbf{78}, 011301 (2007).

\bibitem{Shechtman2015PhaseRetrieval}
Y.~Shechtman, Y.~C. Eldar, O.~Cohen \emph{et~al.}, \enquote{Phase retrieval
  with application to optical imaging: a contemporary overview,}
  {\protect\JournalTitle{IEEE Signal Processing Magazine}} \textbf{32}, 87--109
  (2015).

\bibitem{Stockmar2013Nearfield}
M.~Stockmar, P.~Cloetens, I.~Zanette, \emph{et~al.}, \enquote{Near-field
  ptychography: phase retrieval for inline holography using a structured
  illumination,} {\protect\JournalTitle{Scientific Reports}} \textbf{3}, 1927
  (2013).

\bibitem{Marchesini2003Shrinkwrap}
S.~Marchesini, H.~He, H.~N. Chapman, \emph{et~al.}, \enquote{X-ray image
  reconstruction from a diffraction pattern alone,}
  {\protect\JournalTitle{Physical Review B}} \textbf{68}, 140101(R) (2003).

\bibitem{Chapman2006FemtoCDI}
H.~N. Chapman, A.~Barty, M.~J. Bogan \emph{et~al.}, \enquote{Femtosecond
  diffractive imaging with a soft-{X}-ray free-electron laser,}
  {\protect\JournalTitle{Nature Physics}} \textbf{2}, 839--843 (2006).

\bibitem{Seibert2011Nature}
M.~M. Seibert, T.~Ekeberg, F.~R. N.~C. Maia \emph{et~al.}, \enquote{Single
  mimivirus particles intercepted and imaged with an {X}-ray laser,}
  {\protect\JournalTitle{Nature}} \textbf{470}, 78--81 (2011).

\bibitem{Takayama2021MultipleShotCDI}
Y.~Takayama, K.~Fukuda, M.~Kawashima, \emph{et~al.}, \enquote{Dynamic
  nanoimaging of extended objects via hard x-ray multiple-shot coherent
  diffraction with projection illumination optics,}
  {\protect\JournalTitle{Communications Physics}} \textbf{4}, 48 (2021).

\bibitem{Hoppe1969}
W.~Hoppe, \enquote{Beugung im inhomogenen prim{\"a}rstrahlwellenfeld. {I}.
  {P}rinzip einer phasenmessung von elektronenbeugungsinterferenzen,}
  {\protect\JournalTitle{Acta Crystallographica Section A}} \textbf{25},
  495--501 (1969).

\bibitem{Rodenburg2004APL}
J.~M. Rodenburg and H.~M.~L. Faulkner, \enquote{A phase retrieval algorithm for
  shifting illumination,} {\protect\JournalTitle{Applied Physics Letters}}
  \textbf{85}, 4795--4797 (2004).

\bibitem{Thibault2008Science}
P.~Thibault, M.~Dierolf, A.~Menzel \emph{et~al.}, \enquote{High-resolution
  scanning {X}-ray diffraction microscopy,} {\protect\JournalTitle{Science}}
  \textbf{321}, 379--382 (2008).

\bibitem{Maiden2009UltramicroscopyPIE}
A.~M. Maiden and J.~M. Rodenburg, \enquote{An improved ptychographical phase
  retrieval algorithm for diffractive imaging,}
  {\protect\JournalTitle{Ultramicroscopy}} \textbf{109}, 1256--1262 (2009).

\bibitem{GuizarSicairos2008TTD}
M.~Guizar-Sicairos and J.~R. Fienup, \enquote{Phase retrieval with transverse
  translation diversity: a nonlinear optimization approach,}
  {\protect\JournalTitle{Optics Express}} \textbf{16}, 7264--7278 (2008).

\bibitem{Maiden2017rPIE}
A.~Maiden, D.~Johnson, and P.~Li, \enquote{Further improvements to the
  ptychographical iterative engine,} {\protect\JournalTitle{Optica}}
  \textbf{4}, 736--745 (2017).

\bibitem{Bunk2008Overlap}
O.~Bunk, M.~Dierolf, S.~Kynde, \emph{et~al.}, \enquote{Influence of the overlap
  parameter on the convergence of the ptychographical iterative engine,}
  {\protect\JournalTitle{Ultramicroscopy}} \textbf{108}, 481--487 (2008).

\bibitem{Edo2013PRA}
T.~B. Edo, D.~J. Batey, A.~M. Maiden \emph{et~al.}, \enquote{Sampling in x-ray
  ptychography,} {\protect\JournalTitle{Physical Review A}} \textbf{87}, 053850
  (2013).

\bibitem{Zheng2013FPM}
G.~Zheng, R.~Horstmeyer, and C.~Yang, \enquote{Wide-field, high-resolution
  {F}ourier ptychographic microscopy,} {\protect\JournalTitle{Nature
  Photonics}} \textbf{7}, 739--745 (2013).

\bibitem{Thibault2009Probe}
P.~Thibault, M.~Dierolf, O.~Bunk \emph{et~al.}, \enquote{Probe retrieval in
  ptychographic coherent diffractive imaging,}
  {\protect\JournalTitle{Ultramicroscopy}} \textbf{109}, 338--343 (2009).

\bibitem{Maiden2012Annealing}
A.~M. Maiden, M.~J. Humphry, M.~C. Sarahan, \emph{et~al.}, \enquote{An
  annealing algorithm to correct positioning errors in ptychography,}
  {\protect\JournalTitle{Ultramicroscopy}} \textbf{120}, 64--72 (2012).

\bibitem{Zhang2013Position}
F.~Zhang, I.~Peterson, J.~Vila-Comamala, \emph{et~al.}, \enquote{Translation
  position determination in ptychographic coherent diffraction imaging,}
  {\protect\JournalTitle{Optics Express}} \textbf{21}, 13592--13606 (2013).

\bibitem{Nashed2014Parallel}
Y.~S.~G. Nashed, D.~J. Vine, T.~Peterka \emph{et~al.}, \enquote{Parallel
  ptychographic reconstruction,} {\protect\JournalTitle{Optics Express}}
  \textbf{22}, 32082--32097 (2014).

\bibitem{Marchesini2016SHARP}
S.~Marchesini, H.~Krishnan, B.~J. Daurer, \emph{et~al.}, \enquote{{SHARP}: a
  distributed {GPU}-based ptychographic solver,} {\protect\JournalTitle{Journal
  of Applied Crystallography}} \textbf{49}, 1245--1252 (2016).

\bibitem{Enders2016PtyPy}
B.~Enders and P.~Thibault, \enquote{A computational framework for ptychographic
  reconstructions,} {\protect\JournalTitle{Proceedings of the Royal Society A}}
  \textbf{472}, 20160640 (2016).

\bibitem{FavreNicolin2020PyNX}
V.~Favre-Nicolin, G.~Girard, S.~Leake \emph{et~al.}, \enquote{{PyNX}:
  high-performance computing toolkit for coherent {X}-ray imaging based on
  operators,} {\protect\JournalTitle{Journal of Applied Crystallography}}
  \textbf{53}, 1404--1413 (2020).

\bibitem{Wakonig2020PtychoShelves}
K.~Wakonig, H.-C. Stadler, M.~Odstr{\v{c}}il \emph{et~al.},
  \enquote{{PtychoShelves}, a versatile high-level framework for
  high-performance analysis of ptychographic data,}
  {\protect\JournalTitle{Journal of Applied Crystallography}} \textbf{53},
  574--586 (2020).

\bibitem{Babu2023EdgePtycho}
A.~V. Babu, T.~Zhou, S.~Kandel, \emph{et~al.}, \enquote{Deep learning at the
  edge enables real-time streaming ptychographic imaging,}
  {\protect\JournalTitle{Nature Communications}} \textbf{14}, 7059 (2023).

\bibitem{Sinha2017Optica}
A.~Sinha, J.~Lee, S.~Li, and G.~Barbastathis, \enquote{Lensless computational
  imaging through deep learning,} {\protect\JournalTitle{Optica}} \textbf{4},
  1117--1125 (2017).

\bibitem{Rivenson2018LSA}
Y.~Rivenson, Y.~Zhang, H.~G{\"u}naydin \emph{et~al.}, \enquote{Phase recovery
  and holographic image reconstruction using deep learning in neural networks,}
  {\protect\JournalTitle{Light: Science \& Applications}} \textbf{7}, 17141
  (2018).

\bibitem{Barbastathis2019Optica}
G.~Barbastathis, A.~Ozcan, and G.~Situ, \enquote{On the use of deep learning
  for computational imaging,} {\protect\JournalTitle{Optica}} \textbf{6},
  921--943 (2019).

\bibitem{Cherukara2018SciRep}
M.~J. Cherukara, Y.~S.~G. Nashed, and R.~J. Harder, \enquote{Real-time coherent
  diffraction inversion using deep generative networks,}
  {\protect\JournalTitle{Scientific Reports}} \textbf{8}, 16520 (2018).

\bibitem{Cherukara2020PtychoNN}
M.~J. Cherukara, T.~Zhou, Y.~S.~G. Nashed, \emph{et~al.}, \enquote{{AI}-enabled
  high-resolution scanning coherent diffraction imaging,}
  {\protect\JournalTitle{Applied Physics Letters}} \textbf{117}, 044103 (2020).

\bibitem{Hoidn2023PtychoPINN}
O.~Hoidn, A.~A. Mishra, and A.~Mehta, \enquote{Physics constrained unsupervised
  deep learning for rapid, high resolution scanning coherent diffraction
  reconstruction,} {\protect\JournalTitle{Scientific Reports}} \textbf{13},
  22789 (2023).

\bibitem{Ulyanov2018DIP}
D.~Ulyanov, A.~Vedaldi, and V.~Lempitsky, \enquote{Deep image prior,} in
  \emph{Proceedings of the IEEE Conference on Computer Vision and Pattern
  Recognition (CVPR),}  (2018), pp. 9446--9454.

\bibitem{Metzler2018prDeep}
C.~A. Metzler, P.~Schniter, A.~Veeraraghavan, and R.~G. Baraniuk,
  \enquote{{prDeep}: Robust phase retrieval with a flexible deep network,} in
  \emph{Proceedings of the 35th International Conference on Machine Learning,}
  vol.~80 of \emph{Proceedings of Machine Learning Research} (2018), pp.
  3501--3510.

\bibitem{McCray2025IntegratedNN}
A.~R.~C. McCray, S.~M. Ribet, G.~Varnavides, and C.~Ophus,
  \enquote{Accelerating iterative ptychography with an integrated neural
  network,} {\protect\JournalTitle{Journal of Microscopy}} \textbf{300},
  180--190 (2025).

\bibitem{Sitzmann2020SIREN}
V.~Sitzmann, J.~N.~P. Martel, A.~W. Bergman, \emph{et~al.}, \enquote{Implicit
  neural representations with periodic activation functions,} in \emph{Advances
  in Neural Information Processing Systems,}  vol.~33 (2020), pp. 7462--7473.

\bibitem{Du2024ProbePosition}
M.~Du, T.~Zhou, J.~Deng, \emph{et~al.}, \enquote{Predicting ptychography probe
  positions using single-shot phase retrieval neural network,}
  {\protect\JournalTitle{Optics Express}} \textbf{32}, 36757--36780 (2024).

\bibitem{Gan2024PtychoDV}
W.~Gan, Q.~Zhai, M.~T. McCann, \emph{et~al.}, \enquote{{PtychoDV}: Vision
  transformer-based deep unrolling network for ptychographic image
  reconstruction,} {\protect\JournalTitle{IEEE Open Journal of Signal
  Processing}} \textbf{5}, 539--547 (2024).

\bibitem{Wengrowicz2020DeepSSP}
O.~Wengrowicz, O.~Peleg, T.~Zahavy \emph{et~al.}, \enquote{Deep neural networks
  in single-shot ptychography,} {\protect\JournalTitle{Optics Express}}
  \textbf{28}, 17511--17520 (2020).

\bibitem{Chan2021APR}
H.~Chan, Y.~S.~G. Nashed, S.~Kandel \emph{et~al.}, \enquote{Rapid {3D}
  nanoscale coherent imaging via physics-aware deep learning,}
  {\protect\JournalTitle{Applied Physics Reviews}} \textbf{8}, 021407 (2021).

\bibitem{Yao2022AutoPhaseNN}
Y.~Yao, H.~Chan, S.~K. R.~S. Sankaranarayanan, \emph{et~al.},
  \enquote{{AutoPhaseNN}: unsupervised physics-aware deep learning of {3D}
  nanoscale {Bragg} coherent diffraction imaging,} {\protect\JournalTitle{npj
  Computational Materials}} \textbf{8}, 124 (2022).

\bibitem{Zhang2024FPR}
Z.~Zhang, F.~Wang, Q.~Min, \emph{et~al.}, \enquote{Fourier phase retrieval
  using physics-enhanced deep learning,} {\protect\JournalTitle{Optics
  Letters}} \textbf{49}, 6129--6132 (2024).

\bibitem{Raissi2019PINN}
M.~Raissi, P.~Perdikaris, and G.~E. Karniadakis, \enquote{Physics-informed
  neural networks: a deep learning framework for solving forward and inverse
  problems involving nonlinear partial differential equations,}
  {\protect\JournalTitle{Journal of Computational Physics}} \textbf{378},
  686--707 (2019).

\bibitem{Karniadakis2021PIML}
G.~E. Karniadakis, I.~G. Kevrekidis, L.~Lu \emph{et~al.},
  \enquote{Physics-informed machine learning,} {\protect\JournalTitle{Nature
  Reviews Physics}} \textbf{3}, 422--440 (2021).

\bibitem{Sidorenko2015Optica}
P.~Sidorenko and O.~Cohen, \enquote{Single-shot ptychography,}
  {\protect\JournalTitle{Optica}} \textbf{3}, 9--14 (2016).

\bibitem{Kharitonov2022SciRep}
K.~Kharitonov, M.~Mehrjoo, M.~Ruiz-Lopez, \emph{et~al.}, \enquote{Single-shot
  ptychography at a soft {X-ray} free-electron laser,}
  {\protect\JournalTitle{Scientific Reports}} \textbf{12}, 14430 (2022).

\bibitem{Zhang2016CMI}
F.~Zhang, B.~Chen, G.~R. Morrison, \emph{et~al.}, \enquote{Phase retrieval by
  coherent modulation imaging,} {\protect\JournalTitle{Nature Communications}}
  \textbf{7}, 13367 (2016).

\bibitem{Dong2018CMI}
X.~Dong, X.~Pan, C.~Liu, and J.~Zhu, \enquote{Single shot multi-wavelength
  phase retrieval with coherent modulation imaging,}
  {\protect\JournalTitle{Optics Letters}} \textbf{43}, 1762--1765 (2018).

\bibitem{Paganin2006CoherentXRayOptics}
D.~M. Paganin, \emph{Coherent X-Ray Optics} (Oxford University Press, 2006).
  Appendix B: the Fresnel scaling theorem.

\bibitem{Vong2025GeneralizablePtycho}
A.~Vong, S.~Henke, O.~Hoidn, \emph{et~al.}, \enquote{Towards generalizable deep
  ptychography neural networks,} {\protect\JournalTitle{npj Computational
  Materials}}  (2026). \url{https://doi.org/10.1038/s41524-026-02198-4}.

\bibitem{miao1999extending}
J.~Miao, P.~Charalambous, J.~Kirz, and D.~Sayre, \enquote{Extending the
  methodology of {X-ray} crystallography to allow imaging of micrometre-sized
  non-crystalline specimens,} {\protect\JournalTitle{Nature}} \textbf{400},
  342--344 (1999).

\bibitem{Thibault2012NJPML}
P.~Thibault and M.~Guizar-Sicairos, \enquote{Maximum-likelihood refinement for
  coherent diffractive imaging,} {\protect\JournalTitle{New Journal of
  Physics}} \textbf{14}, 063004 (2012).

\bibitem{Seifert2023PoissonGaussian}
J.~Seifert, Y.~Shao, R.~van Dam, \emph{et~al.}, \enquote{Maximum-likelihood
  estimation in ptychography in the presence of {Poisson}--{Gaussian} noise
  statistics,} {\protect\JournalTitle{Optics Letters}} \textbf{48}, 6027--6030
  (2023).

\bibitem{Kandel2019AD}
S.~Kandel, S.~Maddali, M.~Allain \emph{et~al.}, \enquote{Using automatic
  differentiation as a general framework for ptychographic reconstruction,}
  {\protect\JournalTitle{Optics Express}} \textbf{27}, 18653--18672 (2019).

\bibitem{Du2021Adorym}
M.~Du, S.~Kandel, J.~Deng \emph{et~al.}, \enquote{Adorym: a multi-platform
  generic {X}-ray image reconstruction framework based on automatic
  differentiation,} {\protect\JournalTitle{Optics Express}} \textbf{29},
  10000--10035 (2021).

\bibitem{Chollet2015XPP}
M.~Chollet, R.~Alonso-Mori, M.~Cammarata, \emph{et~al.}, \enquote{The x-ray
  pump-probe instrument at the linac coherent light source,}
  {\protect\JournalTitle{Journal of Synchrotron Radiation}} \textbf{22},
  503--507 (2015).

\bibitem{Deng2022MultiPass}
J.~Deng, Y.~Yao, Y.~Jiang, \emph{et~al.}, \enquote{High-resolution
  ptychographic imaging enabled by high-speed multi-pass scanning,}
  {\protect\JournalTitle{Optics Express}} \textbf{30}, 26027--26042 (2022).

\bibitem{Deng2019Velociprobe}
J.~Deng, C.~Preissner, J.~A. Klug, \emph{et~al.}, \enquote{The velociprobe: An
  ultrafast hard x-ray nanoprobe for high-resolution ptychographic imaging,}
  {\protect\JournalTitle{Review of Scientific Instruments}} \textbf{90}, 083701
  (2019).

\bibitem{Du2025PtyChi}
M.~Du, H.~Ruth, S.~Henke, \emph{et~al.}, \enquote{{Pty-Chi}: A {PyTorch}-based
  modern ptychographic data analysis package,} {\protect\JournalTitle{arXiv}}
  2510.20929 (2025).

\bibitem{Odstrcil2018LSQML}
M.~Odstr{\v{c}}il, A.~Menzel, and M.~Guizar-Sicairos, \enquote{Iterative
  least-squares solver for generalized maximum-likelihood ptychography,}
  {\protect\JournalTitle{Optics Express}} \textbf{26}, 3108--3123 (2018).

\bibitem{gao2023iterative}
Y.~Gao and L.~Cao, \enquote{Iterative projection meets sparsity regularization:
  towards practical single-shot quantitative phase imaging with in-line
  holography,} {\protect\JournalTitle{Light: Advanced Manufacturing}}
  \textbf{4}, 37--53 (2023).

\bibitem{Howells2009Dose}
M.~R. Howells, T.~Beetz, H.~N. Chapman \emph{et~al.}, \enquote{An assessment of
  the resolution limitation due to radiation-damage in x-ray diffraction
  microscopy,} {\protect\JournalTitle{Journal of Electron Spectroscopy and
  Related Phenomena}} \textbf{170}, 4--12 (2009).

\bibitem{Li2021FNO}
Z.~Li, N.~Kovachki, K.~Azizzadenesheli \emph{et~al.}, \enquote{Fourier neural
  operator for parametric partial differential equations,} in
  \emph{International Conference on Learning Representations (ICLR),}  (2021).

\end{thebibliography}

\clearpage
\setcounter{section}{0}\setcounter{figure}{0}\setcounter{equation}{0}
\renewcommand{\thesection}{S\arabic{section}}
\renewcommand{\thefigure}{S\arabic{figure}}
\renewcommand{\theequation}{S\arabic{equation}}
\makeatletter\@ifpackageloaded{hyperref}{\renewcommand{\theHsection}{S\arabic{section}}\renewcommand{\theHfigure}{S\arabic{figure}}\renewcommand{\theHequation}{S\arabic{equation}}}{}\makeatother
\section*{Supplementary Material}

\section{Effective-probe scaled-Fresnel equivalence}

We evaluate the effective-probe convention on the retrieved complex LCLS XPP fields rather than on a synthetic envelope. The calculation uses the retrieved $64\times64$ probe and a representative $64\times64$ object patch centered at the integer part of the scan coordinate for frame 500, $(x,y)=(50,59)$ pixels. Let
\begin{equation}
 C_z(\vec x)=\exp\!\left(\frac{\mathrm{i}\pi|\vec x|^2}{\lambda z}\right),
 \qquad P_{\mathrm{physical}}=P_{\mathrm{eff}}C_z^*.
\end{equation}
The one-step discrete Fresnel propagation of $P_{\mathrm{physical}}O$ multiplies the input by $C_z$ before the Fourier transform. It therefore reduces to the implemented effective-probe map $\mathcal F\{P_{\mathrm{eff}}O\}$, apart from the common propagation prefactor and an output quadratic phase that cancels in intensity. Both intensity calculations use the same centered coordinates, centered-FFT convention, and propagation prefactor; the displayed detector line uses one common Fresnel-peak normalization.

\begin{figure}[!htbp]
  \centering
  \includegraphics[width=\linewidth]{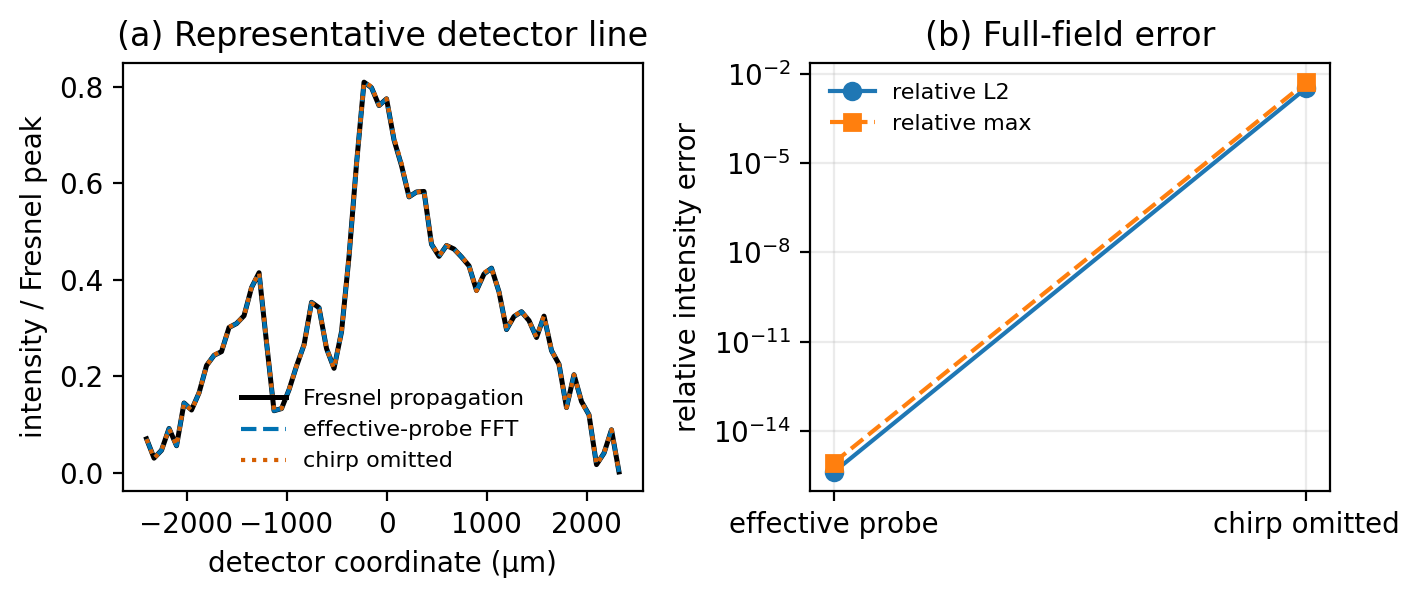}
  \caption{Scaled-Fresnel effective-probe check using the retrieved LCLS XPP probe and a representative complex object patch. (a) Central detector line for one-step Fresnel propagation, the equivalent effective-probe Fourier map, and the deliberately mismatched control with the input chirp omitted. (b) Full-field relative intensity errors. The equivalent curves agree to floating-point precision; the mismatched control separates only in the error panel (b).}
  \label{fig:fresnel_equivalence}
\end{figure}

The calculation uses $\lambda=1.3947576\times10^{-10}$~m, $z=4.147$~m, detector pitch $\Delta u=75~\mu$m, $N=64$, and the supplied rounded object pitch $\Delta x=120.501$~nm. The exact sampling closure $\lambda z/(N\Delta u)$ is $120.501245$~nm; equivalently, the rounded $\Delta x$ gives $\Delta u=75.000153~\mu$m, a relative discrepancy of $2.03\times10^{-6}$. For intensity arrays $I_{\mathrm{F}}$ and $I$, we report
\begin{equation}
 E_2=\frac{\lVert I_{\mathrm{F}}-I\rVert_2}{\lVert I_{\mathrm{F}}\rVert_2},
 \qquad
 E_\infty=\frac{\max|I_{\mathrm{F}}-I|}{\max I_{\mathrm{F}}}.
\end{equation}
The effective-probe Fourier map gives $E_2=4.21\times10^{-16}$ and $E_\infty=8.72\times10^{-16}$. As a negative control, omitting $C_z$ and Fourier-transforming $P_{\mathrm{physical}}O$ gives $E_2=3.36\times10^{-3}$ and $E_\infty=5.29\times10^{-3}$. Figure~\ref{fig:fresnel_equivalence} compares the equivalent and negative-control calculations. This calculation verifies the numerical equivalence between explicit Fresnel propagation and the Fourier model using an effective probe that contains the Fresnel quadratic phase. It does not independently validate the experimental geometry.

\clearpage

\section{LCLS XPP Fourier-domain reference-agreement sensitivity}

\begin{figure}[!htbp]
  \centering
  \includegraphics[width=\linewidth]{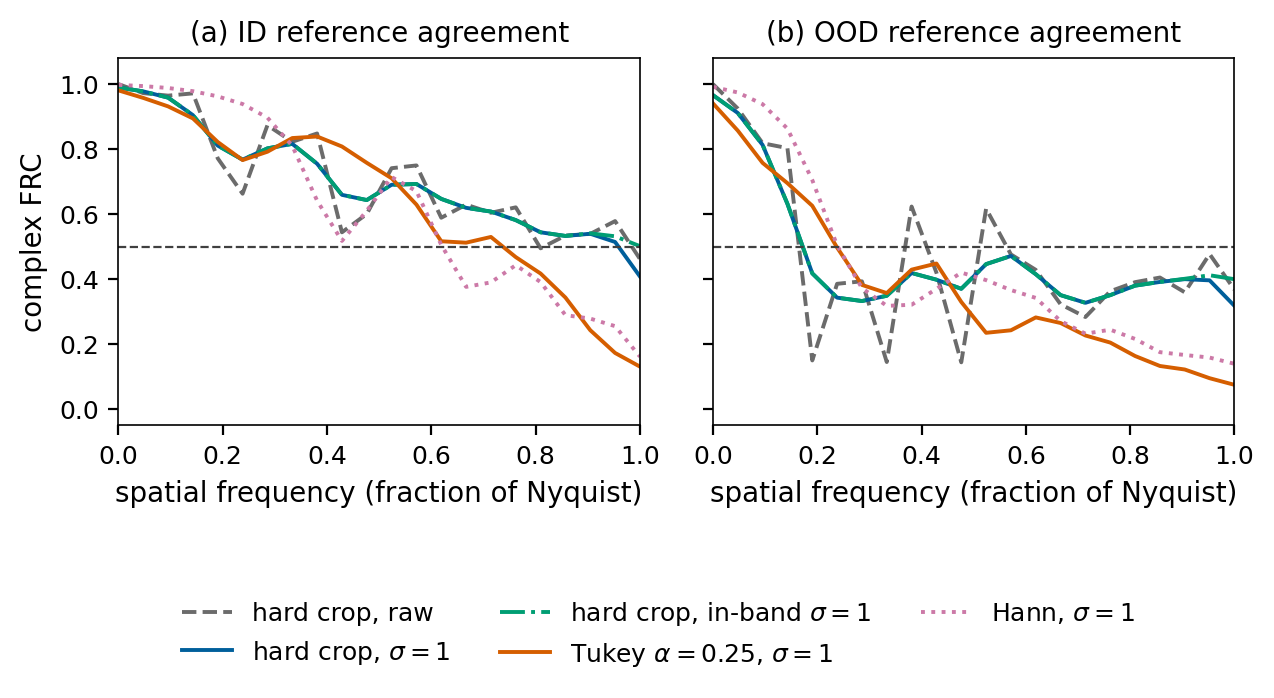}
  \caption{Processing sensitivity of complex Fourier-domain agreement with the ePIE reference over the held-out LCLS XPP test half. (a) In-distribution PtychoPINN. (b) Negative-control PtychoPINN trained on APS data and evaluated on LCLS XPP. Every curve uses the same frozen aligned complex fields; only crop apodization and radial-smoothing order differ. The horizontal 0.5 line is a visual guide, not a resolution threshold. No crossing is interpreted as specimen resolution because the terminal in-distribution behavior is processing-sensitive and ePIE is not independent ground truth.}
  \label{fig:frc_sensitivity}
\end{figure}

Let $U(\vec q)$ and $V(\vec q)$ be the discrete Fourier transforms of the aligned PtychoPINN and ePIE complex fields, and let $\mathcal A_\rho$ contain samples assigned to integer-radius annulus $\rho$. We compute
\begin{equation}
 \mathrm{FRC}(\rho)=
 \frac{\left|\sum_{\vec q\in\mathcal A_\rho}U(\vec q)V^*(\vec q)\right|}
 {\sqrt{\left(\sum_{\vec q\in\mathcal A_\rho}|U(\vec q)|^2\right)
 \left(\sum_{\vec q\in\mathcal A_\rho}|V(\vec q)|^2\right)}} .
\end{equation}
The held-out arrays are center-cropped to a common $42\times42$ square; no support mask or fine registration is applied. Figure~\ref{fig:frc_sensitivity} compares five processing choices on the same frozen aligned arrays: the raw hard-crop curve; the original hard-crop curve Gaussian-smoothed over the full radial sequence with $\sigma=1$ bin before truncation at Nyquist; the hard-crop curve truncated at Nyquist before the same smoothing; and separable Tukey ($\alpha=0.25$) or Hann apodization before the Fourier transform, followed by full-radial $\sigma=1$ smoothing and Nyquist truncation. The original smoothed hard-crop curves reproduce the archived metric arrays exactly.

For the in-distribution field, the first nominal crossing below 0.5 ranges from $0.621$ to $0.959$ of Nyquist among choices that cross, while smoothing only the in-band samples produces no crossing before Nyquist. The raw curve first crosses at $0.808$ and subsequently rises above 0.5. For the cross-facility negative control, the corresponding first-crossing locations span $0.165$--$0.239$ of Nyquist. This sensitivity, together with the small crop and the use of a same-acquisition ePIE reconstruction rather than statistically independent half-data estimates, prevents a threshold crossing from serving as a spatial-resolution estimate. The curves are retained only as frequency-resolved reference-agreement diagnostics.

\end{document}